\documentclass{aa}

\usepackage{graphicx}
\usepackage{txfonts}
\usepackage{amssymb}
\begin{document}

   \title{The treasure behind the haystack:\\ MUSE analysis of five recently discovered globular clusters\thanks{Based 
   on observations collected at the European Southern Observatory under ESO programmes 0103.D-0386(A) and 105.20MY.001 (PI: F. Gran).}}

   % \subtitle{I. Overviewing the $\kappa$-mechanism}

   \author{F. Gran \inst{1,2,3}
          \and
          G. Kordopatis \inst{1}
          \and
          M. Zoccali \inst{2,3}
          \and
          V. Hill \inst{1}
          \and
          I. Saviane \inst{4}
          \and
          C. Navarrete \inst{1,4}
          \and
          A. Rojas-Arriagada \inst{5,3,6,7}
          \and
          \\J. Carballo-Bello \inst{8}
          \and
          J. Hartke \inst{9,10}
          \and
          E. Valenti \inst{11}
          \and
          R. Contreras Ramos \inst{3}
          \and
          M. De Leo \inst{2,3}
          \and
          S. Fabbro \inst{1,12}
          }

   \institute{Universit\'e C\^ote d’Azur, Observatoire de la C\^ote d’Azur, CNRS, Laboratoire Lagrange, Nice, France,
   \email{fgran@oca.eu}
   \and
   Instituto de Astrofísica, Av. Vicuna Mackenna 4860, Santiago, Chile
   \and
   Millennium Institute of Astrophysics, Av. Vicu\~{n}a Mackenna 4860, 82-0436 Macul, Santiago, Chile
   \and
   European Southern Observatory, Alonso de Córdova 3107, Casilla 19001 Santiago, Chile
   \and
   Departamento de F\'isica, Universidad de Santiago de Chile, Av. Victor Jara 3659, Santiago, Chile
   \and
   N\'ucleo Milenio ERIS
   \and
   Center for Interdisciplinary Research in Astrophysics and Space Exploration (CIRAS), Universidad de Santiago de Chile, Santiago, Chile
   \and
   Instituto de Alta Investigaci\'on, Universidad de Tarapac\'a, Casilla 7D, Arica, Chile
   \and
   Finnish Centre for Astronomy with ESO (FINCA), FI-20014 University of Turku, Finland
   \and
   Tuorla Observatory, Department of Physics and Astronomy, FI-20014 University of Turku, Finland
   \and
   European Southern Observatory, Karl Schwarzschild-Strabe 2, 85748 Garching bei München, Germany
   \and
   National Research Council of Canada, Herzberg Astronomy \& Astrophysics Research Centre, 5071 West Saanich Road, Victoria, BC V9E 2E7, Canada
   }
   \date{Received ; accepted }

% \abstract{}{}{}{}{} 
% 5 {} token are mandatory
 
\abstract
{After the second data release of Gaia, the number of new globular cluster candidates has increased importantly. 
However, most of them need to be properly characterised, both spectroscopically and photometrically, by means of radial velocities, metallicities, and deeper photometric observations.}
{Our goal is to provide an independent confirmation of the cluster nature of Gran~4, a recently discovered globular cluster, with follow-up spectroscopic observations.
The derived radial velocity for individual stars, coupled with proper motions, allows us to isolate cluster members from field stars, while the analysis of their spectra allows us to derive metallicities.
By including in the analysis the recently confirmed clusters Gran~1, 2, 3, and 5, we aim to completely characterise the sample presented in \citealt{gran2022}.}
{Using Gaia DR3 and VVV catalogue data and MUSE@VLT observations, we performed a selection of cluster members based on their proper motions, radial velocities and their position in colour-magnitude diagrams.
Furthermore, full spectral synthesis was performed on the cluster members, extracting surface parameters and metallicity from MUSE spectra.
Finally, a completeness estimation was performed on the total globular cluster population of the Milky Way.}
{We confirm the nature of Gran~4, a newly discovered globular cluster behind the Galactic bulge, 
with a mean radial velocity of ${\rm RV} = -265.28 \pm 3.92$ km s$^{-1}$ and a mean metallicity of ${\rm [Fe/H] = -1.72 \pm 0.32}$ dex.
Additionally, independent measurements of the metallicities were derived for Gran~1, 2, 3, and 5.
We also revise the observational lower mass limit for a globular cluster to survive in the bulge/disk environment.
We estimate that ~$\sim 12-26$ globular clusters have still to be discovered on the other side of the Galaxy (i.e., behind the bulge/bar/disk), up to 20 kpc.}
{}

\keywords{Surveys -- Stars: kinematics and dynamics -- Galaxy: bulge -- Galaxy: halo -- globular clusters: general -- Proper motions}

\maketitle
%
%-------------------------------------------------------------------

\section{Introduction}
\label{sec:intro}

Globular clusters (GCs) are dense agglomerations of gravitationally bound stars formed roughly at the same epoch and constitute an essential part of galaxies over the whole mass range \citep{brodie2006}.
Although the details of the GC formation process are not yet completely understood \citep{renzini2017,forbes2018}, 
significant progress has been made during the last years to connect high-redshift observations \citep{vanzella2017} with the Milky Way (MW) GC population \citep{gratton2004}.

Historically, GCs have been used as a crucial stellar laboratory given that distances, masses and approximate ages can be derived from their stellar populations \citep{bastian2018}.
Most of our understanding of stellar evolution comes from the analysis of star clusters \citep[either GCs or open clusters, OCs;][and references therein]{gratton2019},
in which the European Space Mission Gaia \citep{gaia} satellite has made a revolutionary contribution to our overall 
understanding of the MW cluster population \citep{bragaglia2018, choi2018, cantat-gaudin2020b, cantat-gaudin2022}.

Given their old ages (mostly $\gtrsim 10$ Gyr), GCs are considered fossil records of the evolution of their host galaxy \citep{harris1979, recioblanco2018},
particularly \cite{martinfranch2009}, \cite{forbes2010} and \cite{leaman2013} first used the MW GC age-metallicity relation to discriminate between accreted and in-situ objects.
More recently, GCs have been used to identify merging events in the MW mass assembly history \citep[][and references therein]{kruijssen2019a, kruijssen2019b, myeong2019, massari2019, callingham2022, hammer2023}.
Nonetheless, \cite{pagnini2022} emphasise the presence of a significant overlap in the kinematic space between potentially accreted and in-situ objects.

The last few years have seen an important increase in the discovery of OCs \citep{castro-ginard2018, cantat-gaudin2019, castro-ginard2020, he2021, hunt2021, castro-ginard2022, he2022, he2023, hunt2023} and GCs \citep{koposov2017, gran2019, garro2020, huang2021, garro2022a, garro2022b, gran2022}, thanks to the proper motions (PMs) measured by the Gaia satellite.
Some new clusters were also discovered towards the Galactic bulge, even if the stellar extinction and crowding place a veil on the Gaia optical observations.
However, most of these discoveries still need to be confirmed with radial velocities (RVs), deeper photometric observations, or even PM analysis.
As shown in \cite{cantat-gaudin2020a} and \cite{gran2019}, the study of PMs has ruled out most of the cluster candidates proposed in the literature. 
To date, the total number of confirmed GCs in the MW is around 170 \citep{baumgardt2021, vasiliev2021}.

\citet{gran2022} identified five new GCs towards the MW bulge, namely Gran~1, 2, 3, 4, and 5.
They showed clustered PMs and well-defined evolutionary sequences in the colour-magnitude diagram (CMD).
All of them were followed up spectroscopically to derive RV and metallicity measurements, 
unambiguously confirming the nature of Gran~1, 2, 3 and 5 (those with available data at the time of publication) in \cite{gran2022}.
Positional, structural, and orbital parameters were also derived, showing that some of them lie on the other side of the bulge.
The spectra for Gran~4 were obtained after the publication of that paper, and this is the reason we discuss them here. 
More recently, \cite{pace2023} independently confirmed the GC nature of two of those clusters (Gran~3 and 4) using archival data and a follow-up spectroscopic campaign.

This article presents the measurements of RV and metallicity for stars in Gran~4 using MUSE data. 
The mean RV derived agrees well with recent studies. 
Our analysis complements the \cite{pace2023} work, given that we precisely point at the inner 1 arcmin of each cluster at once.
Additionally, the independently derived metallicity measurements for all the GCs presented in \cite{gran2022}, 
using low-resolution spectra that cover the complete optical regime ($\sim 5000$ to $\sim 9000$ \r{A}.) is included.

The paper is organised as follows. 
Section~\ref{sec:data_obs} lists the used catalogues and surveys in this work,
Section~\ref{sec:muse} describes the analysis of the low-resolution spectra acquired for this project and presents the RV of the newly discovered Gran~4.
Section~\ref{sec:completeness_mass} elaborates on the observed low-mass limit of MW GCs according to their environment and provides an estimated number of missing GCs on the far side of the Galaxy (behind the Galactic bulge).
Finally, Section~\ref{sec:discussionconclusion} presents a discussion on the topic, the conclusions and prospects for this field.

%--------------------------------------------------------------------
\section{Data and observations}
\label{sec:data_obs}

We used the Gaia Data Release 3 \citep[DR3,][]{gaia, gaiadr3} main catalogue to identify cluster members, extract optical magnitudes (Gaia ${\rm G}$, ${\rm BP}$, and ${\rm RP}$), 
and proper motions (${\rm \mu_\alpha \cos{\delta}}$ and ${\rm \mu_\delta}$).
We recall that \cite{gran2019} initially used the Gaia DR2 \citep{gaiadr2} to discover Gran~1, 2, 3, 4, and 5, which we now update to the DR3 version for the present analysis.

In addition, deep near-IR J,K$_{\rm s}$ photometry from the VISTA Variables in the V\'ia L\'actea survey \citep[hereafter VVV;][]{vvv} 
was used to confirm the presence of stellar overdensities at the cluster centres and complement the optical photometry from Gaia. 
For VVV, we used the public catalogues provided in \cite{surot2019}, in which point-spread function (PSF) photometry was extracted from stacked ${\rm J}$ and ${\rm K_s}$ images.
The stacking process resulted in photometry that goes deeper than other similar catalogues \citep[e.g.;][]{contrerasramos2017, virac} at the price of not including near-IR proper motions, nor light curves, for 
the faintest stars.

Finally, two follow-up observing campaigns with the Multi Unit Spectroscopic Explorer \citep[MUSE;][R$\sim$1900-3700]{bacon2010} mounted at the VLT-UT4 (Cerro Paranal, Chile)
during ESO periods P103 and P105 were awarded to point at the discovered clusters. 
The MUSE integral field unit (IFU) allows us to observe a large number of stars (270-640 objects) in each field with a single pointing using the Wide Field Mode (WFM, one sq arcmin field-of-view).
Additionally, we used the GALACSI adaptive optics system \citep{stuik2006, strobele2012, hartke2020} to improve the spatial resolution, which is crucial for these crowded objects. 
Standard one-hour observing blocks (OBs) were executed between July 2019 and September 2021 with a typical image quality of $\sim 0.8-0.9$ arcsec per exposure.
The processing of raw data was done by the automatic MUSE pipeline \citep{weilbacher2020}, using standard calibrations, and the reduced datacubes were retrieved from the ESO User Portal.
Note that MUSE wavelength values are calibrated using arc line wavelengths in standard air and a conversion must be done when transforming it into the vacuum reference frame \citep{weilbacher2020}.

\section{Analysis of the MUSE datacubes and confirmation of the GC Gran~4}
\label{sec:muse}

\subsection{Extraction and Analysis of MUSE spectra}
\label{sec:analysis1}

We follow the procedure described in \cite{gran2022} to extract stellar spectra from MUSE cubes; however, a few key modifications were made to the analysis pipeline.
The derivation of stellar atmospheric parameters was improved to use the entire extracted spectrum from the MUSE cubes.
We summarise here the extraction procedure and spectral analysis.

The stellar sources were selected in the {\it I}-band image, which was obtained by convolving the MUSE cubes with the wavelength transmission function of the filter.
The same procedure was performed to obtain {\it V}- and {\it R}-band images that will be used later in the analysis.
To derive RVs for the observed stars, flux for all the sources 5$\sigma$ above the background level were extracted in the {\it I}-band images, for each slice of the MUSE cubes. 
The primary motivation of this threshold is to remove low signal-to-noise ratio (SNR) stars that will be present in the field of view with no reliable spectra.
This task was performed using the associated {\tt Astropy}\footnote{\url{https://www.astropy.org/}} package, 
{\tt photutils}\footnote{\url{https://photutils.readthedocs.io/en/stable/}} \citep{astropy1, astropy2, astropy3, photutils}.
After all the available cube layers were extracted, we created individual stellar spectra from $\sim 5000$ to $\sim 9000$ \r{A}.
The entire wavelength range was used in the derivation of both RV and stellar atmospheric parameters.

RVs for all the stars were calculated using the cross-correlation function implemented in the python package {\tt doppler} \citep{nidever2021}.
The package fits RV, $T_{\rm eff}$, $\log\ g$, and ${\rm [Fe/H]}$ at the same time, using a pre-computed grid of synthetic spectra. 
% It is based on the data-driven code The Cannon \citep{ness2015} to derive the observed spectral features, a procedure that was repeated 100 times in a MCMC ensemble \citep{emcee} to extract parameter uncertainties.
It is based on the data-driven code The Cannon \citep{ness2015} to derive the observed spectral features, whose phase-space was explored in a MCMC ensemble \citep{emcee} to extract parameter uncertainties.
Given the model flexibility and the large parameter space covered by the synthetic grid, horizontal branch (HB) stars with prominent Hydrogen lines were also fitted without major issues.

In order to ensure a proper calibration of the RVs and stellar atmospheric parameters derived by {\tt doppler}, we used the AMBRE grid of synthetic spectra \citep{ambre, delaverny2013}. 
Given that our science target is only GC giant stars, we limit the grid to a sub-sample of 456 spectra.
We choose the grid limits based on the expected effective temperatures (${\rm 4250 \leq T_{eff}\ (K) \leq 5250}$, in steps of 250 K), 
surface gravities (${\rm 0 \leq log\ g\ (dex) \leq 3}$, in steps of 0.5 dex),
metallicities (${\rm -3.0 \leq [Fe/H]\ (dex) \leq -1}$, in steps of 0.5 dex), 
and $\alpha$-elements enhancement (${\rm 0.0 \leq [\alpha/Fe]\ (dex) \leq 0.4}$, in steps of 0.2 dex) of red giant stars in other GCs.
Stars outside the cluster RGB, such as foreground disk stars, bulge dwarfs, or GC HB stars will present reliable RVs; however, their atmospheric parameters will not be considered here.

\begin{figure*}
\centering
\includegraphics[scale=0.46]{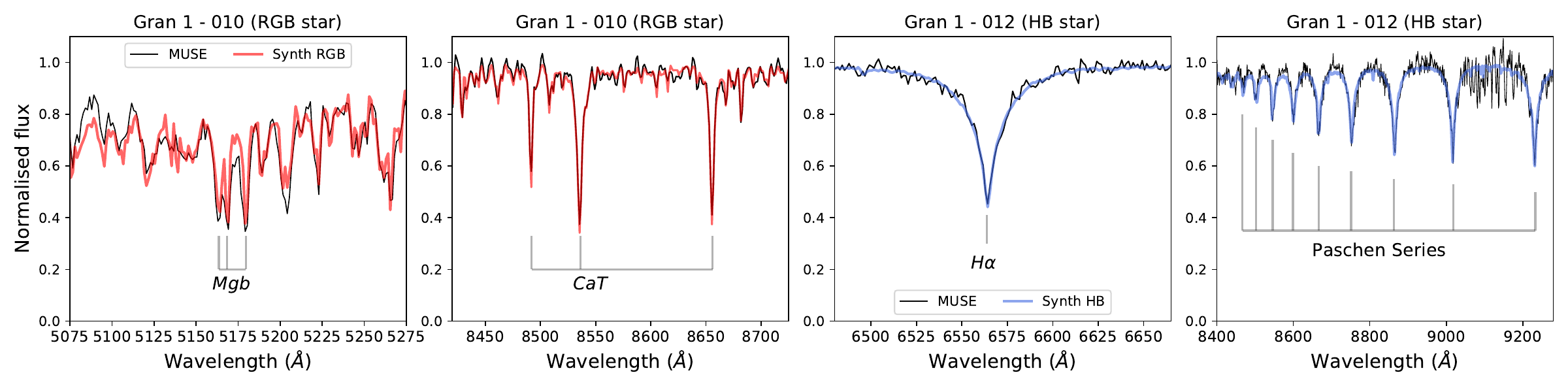}
\caption{Extracted MUSE spectra and associated models for different parts of the spectrum of two high-SNR (>110) stars: Gran~1 stars 010 and 012. In each panel, features in each region are highlighted. 
{\bf Two left panels for RGB star, and two right panels for HB star}: Mgb around $\sim$5175 \AA, H$\alpha$ (6562 \AA), CaT between $\sim8400-8700$ \AA, and the Paschen Hydrogen series starting at $\sim8450$ \AA.}
\label{fig:spectra}
\end{figure*}

Overall, the derived RV, temperature, gravity and metallicity uncertainties were compatible with previous studies \citep{wang2022}, for a given SNR and magnitude.
Mean uncertainty values of the order of $10$ km s$^{-1}$, $100$ K, $0.35$ dex, and $0.35$ dex were found for RVs, temperatures, surface gravities, and metallicities.
Figure~\ref{fig:spectra} shows the spectra of a typical high-SNR giant star of our sample, with emphasis on the Magnesium b triplet (Mgb $\lambda$ 5167, 5172, 5183) 
and Calcium Triplet (CaT; $\lambda$ 8498, 8542, 8662) areas, as well as an HB star with prominent Hydrogen spectral features.
Normalisation was achieved by {\tt doppler} using a 6$^{\rm th}$ order polynomial applied to the continuum pixels (i.e., percentile 90) after a 5-$\sigma$ clipping.
Table~\ref{table:atmos_params} contains all the coordinates and calibrated atmospheric parameters 
derived for each member star during this analysis (i.e., stars within the box in Figures~\ref{fig:triplot} and \ref{fig:triplots}).
Complete derived parameters for all clusters can be found in Appendix~\ref{sec:appendix}, Table~\ref{tbl:allstars1} through Table~\ref{tbl:allstars5}.
Overall, we report here a more accurate measurement of the cluster parameters, lowering the error bars by a factor of $\sim2$ in comparison with \cite{gran2022}.
This change is directly related to the usage of the entire spectra instead of relying on the CaT region alone and its calibration to metallicity.

\begin{table*}
\caption{Stellar atmospheric parameters for MUSE selected Gran~1 members. 
Star ID, coordinates, temperature, surface gravity, metallicity, RV and mean SNR are presented for observed GC members.
Only the first few rows are shown here, however, complete tables can be found in Table~\ref{tbl:allstars1} through \ref{tbl:allstars5} within Appendix~\ref{sec:appendix}.}
\label{table:atmos_params}      
\centering
\tiny
\begin{tabular}{c c c c c c c c}
\hline\hline
GC - ID & RA & Dec & $T_{\rm eff}$ & $\log\ g$ & ${\rm [Fe/H]}$ & RV & SNR\\
        & (deg) & (deg) & (K) & (dex) & (dex) & (km s$^{-1}$) & \\
\hline
Gran 01 - 008 & 269.64794 & -32.02559 & 4362 & 1.2 & -1.17 & 68.25 & 133 \\
Gran 01 - 017 & 269.64650 & -32.02449 & 4659 & 1.4 & -1.11 & 69.33 & 174 \\
Gran 01 - 021 & 269.65005 & -32.02420 & 4358 & 1.1 & -1.14 & 75.04 & 241 \\
Gran 01 - 022 & 269.65615 & -32.02395 & 4611 & 1.4 & -1.05 & 68.81 & 222 \\
\hline                  
\end{tabular}
\end{table*}

% In parallel, our pipeline extracted instrumental VRI magnitudes for all the stars detected 3$\sigma$ above the background in the I-band image. 
Using {\tt TOPCAT} \citep{topcat}, {\it VRI}, Gaia DR3 and VVV data were added to the RV catalogue per cluster. 
This resulted in a catalogue of stars with measured tangential and line-of-sight (LOS) velocities (both PMs and RVs).
We then selected cluster members based on the most prominent group of stars with coherent RVs, PMs and metallicities.
We filter field stars using three times the RV dispersion ($\sigma_{\rm RV}$) indicated in Table~\ref{table:granes} as a threshold.
The RV cuts implemented here (3$\sigma_{\rm RV}$), agree with the maximum of observed dispersion in other MW GCs of $\leq 20$ km s$^{-1}$ \citep{watkins2015}.
Metallicity selection constraints were flexible enough to account for their derived uncertainties (mean of $\sim 0.35$ dex), usually including outliers in the first selection.
Subsequently, we iteratively reduce the selection limits, filtering field stars by their position in the Kiel diagram.

We confirmed the cluster nature of the five GCs analysed through this study.
The final parameters for all the clusters presented in \cite{gran2022} are listed in Table~\ref{table:granes}.
Figure~\ref{fig:triplot} shows the derived RVs, metallicities, and Kiel diagrams (T$_{\rm eff}$-$\log\ g$) for Gran~1 and 4.
For completeness, in Appendix~\ref{sec:appendix} the figures for the other three clusters (Gran~2, 3, and 5) can be found. 
In the case of Gran~1, RVs and metallicities are perfectly clumped in that plane, making the selection process straightforward. 
All the cluster stars belong to the RGB phase, for which the surface parameters are well calibrated, following the correct metallicity slope in the Kiel diagram.
On the other hand, Gran~4 members are comprised of RGB, HB and possibly some sub-giant branch (SGB) or blue straggler (BS) stars, 
artificially spreading the metallicity values into a much wider range than expected for a mono-metallic GC \citep{bailin2019, bailin2022}.
We double-check our metallicity dispersions (${\rm \sigma_{\rm [Fe/H]}}$) with the ones derived in \cite{husser2020} for other MW GCs observed with MUSE, 
and for all but one cluster (Gran~4) their dispersion agrees with a mono-metallic stellar population.
The larger ${\rm \sigma_{\rm [Fe/H]}}$ in Gran 4 can be explained by the lower SNR achieved for the cluster members, located at $\sim20$ kpc, 
and also the relatively low number of measured RGB stars in the instrument field-of-view (15 RGB + 9 HB/SGB/BS stars).

As we mentioned earlier, we calibrated our AMBRE grid targeting only GC RGB stars, leaving other evolutionary stages without proper calibration.
The calibration is evident when comparing the RGB sequence of the cluster with the 10 Gyr cluster mean metallicity PARSEC \citep{bressan2012, marigo2013} isochrones.
For this reason, the surface parameters derived from HB or SGB/BS stars (red crosses) are affected by more significant errors, as shown in Figure~\ref{fig:triplot}.
Nonetheless, when selecting cluster RGB stars (green dots and hatched histogram), the spread is minor, allowing us to assign
a mean metallicity of ${\rm [Fe/H] \sim -1.7}$ to Gran~4, in perfect agreement with the results by \cite{pace2023}, that will be discussed in Section~\ref{sec:discussionconclusion}.

\begin{table*}
\caption{Structural and dynamical parameters derived for the analysed GCs. 
Name, coordinates (equatorial and Galactic), distance, mean RV, cluster observational velocity dispersion, median metallicity, mean PMs and the number of cluster member stars observed are included.}
\label{table:granes}
\centering
\tiny
\begin{tabular}{c c c c c c c c c c c}
\hline\hline
GC & RA & Dec & $\ell$ & b & $d_{\rm LOS}$ & RV $\pm\ \sigma_{\rm RV}$ & ${\rm [Fe/H]} \pm\ {\rm \sigma_{\rm [Fe/H]}}$ & $\mu_\alpha \cos{\delta}$ & $\mu_\delta$ & N$_{\rm stars}$ \\
   & (deg) & (deg) & (deg) & (deg) & (kpc) & (km s$^{-1}$) & (dex) & (mas yr$^{-1}$) & (mas yr$^{-1}$) & \\
\hline
Gran~1   &  269.651 & --32.020 &  --1.233 & --3.977 &  7.94 &    76.98 $\pm$ 3.62 & --1.13 $\pm$  0.06 & --8.10 &  --8.01 & 43 \\
Gran~2   &  257.890 & --24.849 &  --0.771 &   8.587 & 16.60 &    61.24 $\pm$ 2.70 & --1.46 $\pm$  0.13 &   0.19 &  --2.57 & 28 \\
Gran~3   &  256.256 & --35.496 & --10.244 &   3.424 & 10.50 &    91.57 $\pm$ 5.97 & --1.63 $\pm$  0.14 & --3.74 &    0.71 & 33 \\
Gran~4   &  278.113 & --23.114 &   10.198 & --6.388 & 21.90 & --265.28 $\pm$ 3.92 & --1.72 $\pm$  0.32 &   0.51 &  --3.51 & 26 \\
Gran~5   &  267.228 & --24.170 &    4.459 &   1.838 &  4.47 &  --59.19 $\pm$ 4.93 & --1.02 $\pm$  0.11 & --5.32 &  --9.20 & 42 \\
\hline
\end{tabular}
\end{table*}

\begin{figure*}
\centering
\includegraphics[scale=0.45]{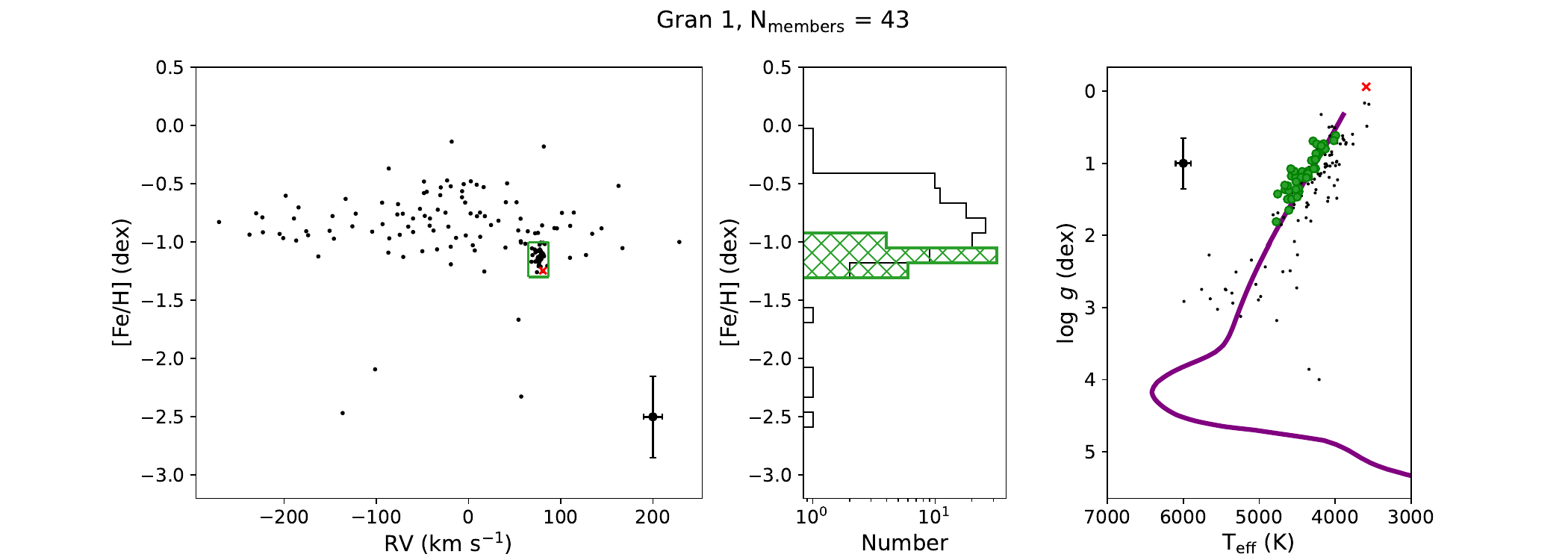}
\includegraphics[scale=0.45]{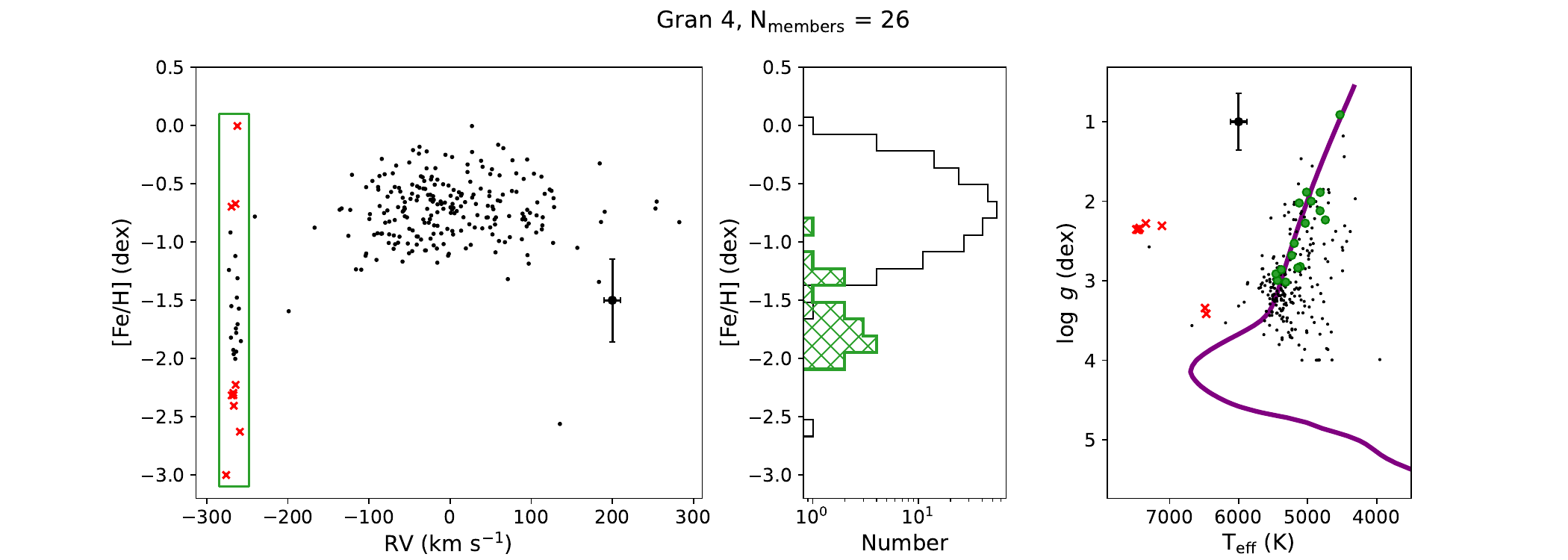}
\caption{{\bf Left panels}: RV-[Fe/H] plane for all the MUSE extracted stars in the Gran~1 ({\bf upper row}) and Gran~4 ({\bf lower row}) fields. 
The box was drawn to select the cluster members using the individual RV and metallicity values. 
Red crosses represent stars outside the calibration parameter space, i.e., stars in the RGB tip or HB/SGB stars, 
which their RV coincides with the one derived for the cluster, but with no reliable stellar atmospheric parameter determination.
{\bf Middle panels}: Metallicity histogram for cluster selected stars. 
The highlighted hatched-filled histogram contains the cluster members with reliable metallicity measurements within the box. 
{\bf Right panels}: Kiel diagram for the field and cluster stars in grey and green, respectively. 
A PARSEC isochrone was drawn using the same derived metallicity with an age of 10 Gyr.
Number of members considers all the valid RV cluster stars, i.e., members within the highlighted box.
The errorbar located in each panel represents the mean uncertainty for each corresponding parameter.}
\label{fig:triplot}
\end{figure*}

\subsection{Confirmation of Gran~4}
\label{sec:analysis2}

\begin{figure*}
\centering
\includegraphics[scale=0.5]{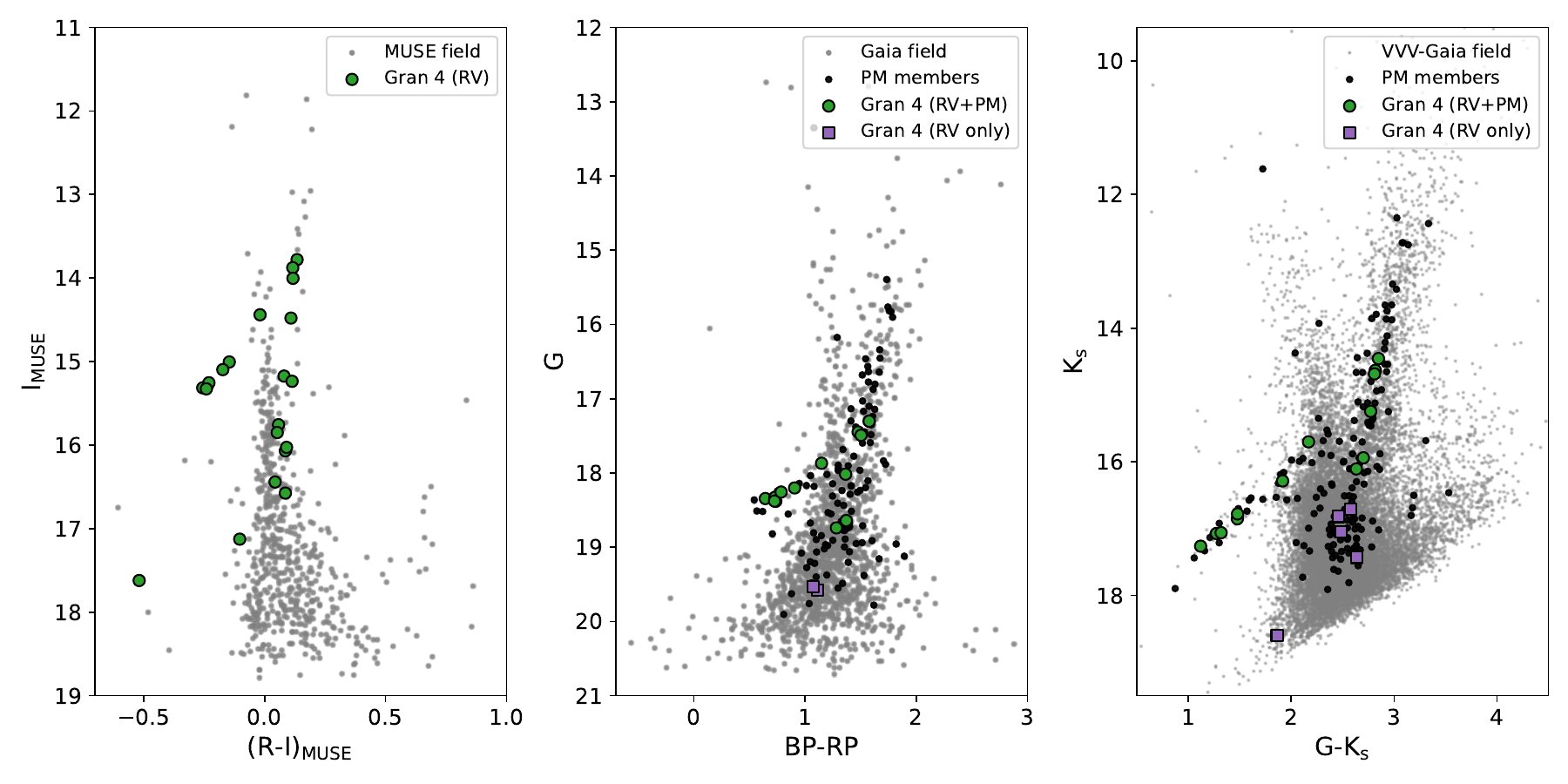}
\caption{CMDs of Gran~4 using different filters and catalogues.
{\bf Left panel}: MUSE CMD of the cluster, containing field stars in the MUSE field-of-view (background points) and cluster members selected by their RVs (green circles).
{\bf Middle panel}: Optical Gaia CMD comprising 2 arcmin within the cluster centre. 
Field stars are showed as background grey points, PM-cleaned members are showed in black points, RV and PM-selected cluster stars as green circles, 
while RV only members (i.e., faint to have PM measurements) are showed in purple squares.
{\bf Right panel}: Same as middle panel with an optical-near-IR CMD.
Note that, for some stars, especially the faint-end, there are only measured ${\rm BP}$ nor ${\rm RP}$-band magnitudes, resulting in a lower number of cluster stars in the CMDs.}
\label{fig:cmd}
\end{figure*}

Given the well-defined sequences in the CMD of Gran~4, see Figure~\ref{fig:cmd}, and \cite{gran2022}, especially the extended HB and its low metallicity and RV, 
well outside the distribution of field stars, the identification of cluster members was straightforward.
We used the 26 isolated spectroscopic members of Gran~4 and matched their coordinates (1 arcsec radius) into the Gaia DR3 catalogue.
In total, 14 stars agree with the cluster RV and PM centroids.
% Those bonafide members were used to derive the mean RV and PM for the cluster.
Considering individual measurement errors, we derive a weighted mean RV and metallicity of --265.28 $\pm$ 3.92 km s$^{-1}$ and --1.72 $\pm$ 0.32 dex, respectively, using 1$\sigma$ uncertainties.
These values agree with a mono-metallic stellar population, for which we can confirm that Gran~4 is indeed a real GC. 
Figure~\ref{fig:cmd} shows the MUSE and Gaia-VVV CMDs for this cluster decontaminated from field stars by means of both PMs and RVs using optical and near-IR filters.

\section{On the Milky Way globular cluster completeness and the initial mass limit of bulge globular clusters}
\label{sec:completeness_mass}

\subsection{Globular cluster completeness behind the Galactic bulge/plane}
\label{sec:GC_completeness}

Recently, multiple efforts have been made to estimate the total number of GCs bound to the Galaxy and, consequently, the number of clusters still to be discovered within the MW.
% Usually, the main focus is given to the distant halo clusters \citep[][and references therein]{contenta2017, webb2021}, assuming completeness of the GC population in the inner Galaxy ($20 < R < 50$ kpc).
Usually, the main focus is given to the distant halo clusters \citep[][and references therein]{contenta2017, webb2021}, 
assuming completeness of the GC population within Galactocentric distances confined to $R \leq 20 - 50$ kpc.
Only a few studies have questioned the completeness of the number of clusters on the other side of the Galaxy \citep{ryu2018} or presented the detection and survival issues of GCs buried in the bulge \citep{minniti2017, minniti2021}.

The clusters presented in this work and other recently confirmed GCs in the inner MW \citep{garro2020, pace2023} challenge the assumption of 100$\%$ completeness towards the Galactic bulge.
In what follows, we will qualitatively estimate the number of missing GCs within Galactocentric distances smaller than 20 kpc and then combine that number with previous halo predictions.
By adding Gran~4, RLGC~1 and 2, and Garro~1 \citep{ryu2018, garro2020} to the compilation by \citet[][March 2023 update]{baumgardt2021}, 
we can give first-order estimates of the missing GCs at the far side of the Galaxy. 
To do this, we consider and adopt the Galactic bar as a symmetry axis and assume that GCs are homogeneously distributed within the MW and that a Poisson distribution can model the number of GCs. 
This assumption is a first-order approximation of the observed GC distribution that exhibits a more complex structure \citep{arakelyan2018}.

\begin{figure}
\centering
\includegraphics[scale=0.55]{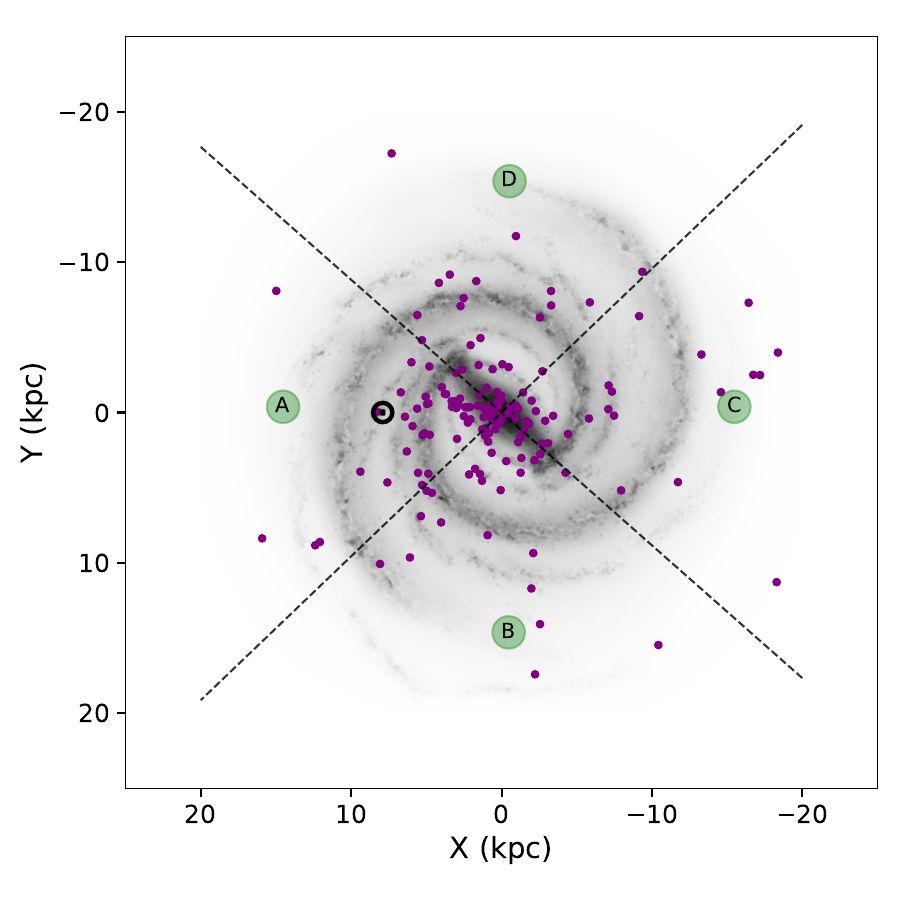}
\caption{Face-on artistic representation of the MW within the inner 20 kpc.%\protect\footnotemark. 
The Sun is at (8, 0) kpc and the known GCs are indicated with $\odot$ and purple filled circles, respectively.
Reference lines are drawn to show the analysed areas (A, B, C and D), considering the Galactic bar as a symmetry axis (i.e., dashed line from top left to down right).}
\label{fig:completeness}
\end{figure}

Figure~\ref{fig:completeness} shows the adopted division to analyse the completeness of the GC population in four regions 
with an artistic representation of the Galaxy in the background\footnote{Image provided in the python package {\tt MW-plot}. Documentation can be found in \protect\url{https://milkyway-plot.readthedocs.io}.}. 
In each of them (A, B, C and D), there are 58, 24, 31 and 31 known GCs, respectively.
In total, 144 GCs are located at Galactocentric distances smaller than 20 kpc.
As expected, a higher number of known GCs are located near the Sun in sector A. 
While roughly the same number of GCs are observed within Poisson errors (i.e., $\sqrt{N}$) in sectors B and D.

Assuming that the existing GCs are distributed isotropically, we can estimate the number of missing clusters in sector C by comparing it with the expected number of clusters observed on the near side of the MW.
For both areas to be statistically similar (1$\sigma$ errorbars), the number of missing clusters should be more than 12 and less than 26.
Given the extensive observations on our side of the Galaxy (i.e., shaded area A), we can almost completely rule out the possibility of number variations in that sector (i.e., undiscovered GC near the Sun).

We repeated the above exercise by (1) counting only GCs currently located close to the plane (${\rm |Z|} < 1.5$ kpc) and (2) limiting $Z_{\rm max} \leq$ 3.5 kpc. 
The results give a similar number of $\sim$11-22 missing GCs.
Adding previous estimations of missing GCs in the halo \citep{contenta2017, webb2021}, 
the combined number of undiscovered clusters reaches up to a mean value of $\sim$22 GCs ((12+26)/2 = 19 behind the Galactic bulge/bar/disk and 3 in the halo).

The number presented above represents a first approximation of the actual value, certainly affected by the violent dynamical processes dominating the early build-up history of the MW. 
In the more recent past of our Galaxy lifetime, only passages through the bar, the bulge and the disk might significantly affect the chance of survival of a GC.

% \footnotetext{Image provided in the python package {\tt MW-plot}. Documentation can be found in \url{https://milkyway-plot.readthedocs.io}.}

\subsection{Observational globular cluster mass limit across the Galaxy}
\label{sec:GC_mass}

Also relevant for the present discussion is the minimum cluster mass that a GC should have to survive and be detected nowadays.
Obviously, the observed GC mass function varies dramatically from the inner MW to the distant halo. 
With only a few GCs missing in the halo, the detection threshold was pushed down to $\sim 10^3 M_\odot$ \citep[see ][]{am4, pal1, whiting1}. 
However, this is one order of magnitude smaller than the mass of the smallest GC detected towards the inner Galaxy.

Complex dynamical processes that occur within the Galaxy, including but not limited to two-body relaxation, tidal truncation, disk/bulge shocks, cluster mass loss, 
dynamical friction and changes in the Galactic potential, create this natural inner-to-outer mass difference \citep[see][and references therein]{murali1997, baumgardt2003, kruijssen2012, carlberg2022, ishchenko2023}. 
Most of the processes listed above occur when GCs are embedded in a dense environment or pass through the disk/bulge along their orbits. 
Recent dynamical analyses of GCs have been performed taking these processes carefully into account \citep[e.g.,][]{baumgardt2018, hughes2020, ferrone2023}, 
therefore, we can translate their results by searching for a lower limit on the GC bulge population.

\begin{figure*}
\centering
\includegraphics[scale=0.6]{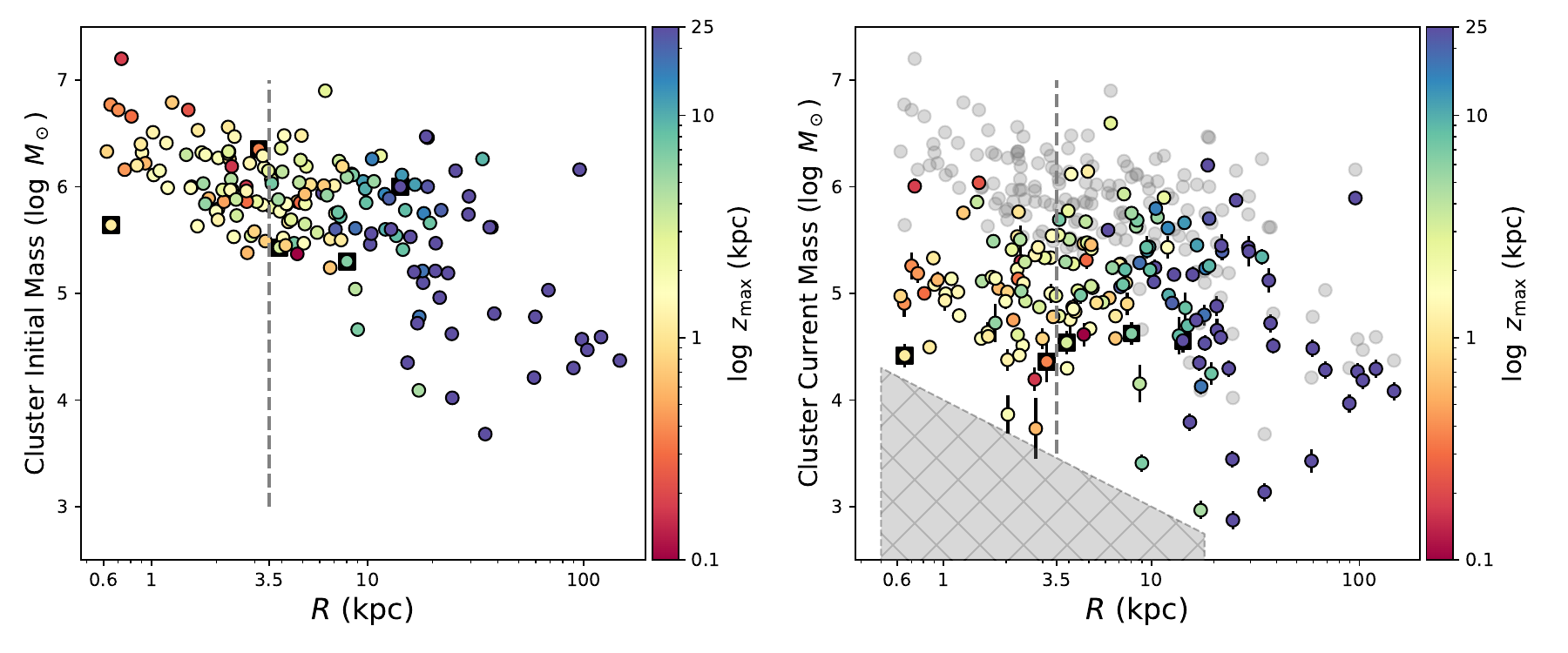}
\caption{
Initial ({\bf left panel}) and current ({\bf right panel}) GC masses as a function of spherical Galactocentric radius ($R$), colour-codded by their $Z_{\rm max}$.
The vertical line at 3.5 kpc is the estimated radius of the Galactic bulge. 
The grey hatched area in the right panel highlights the region where no GC has been discovered, 
implying an observational limit on the mass and orbital configuration below which GCs are more efficiently disrupted by the dynamical processes in the MW.
The clusters analysed in this work are highlighted by squares in both panels. 
They are, in order of increasing $R$: Gran~1, 5, 3, 2, and 4.
The {\bf right panel} also contains the initial cluster masses derived by \protect\cite{baumgardt2019} as grey background points, as a reference for the initial-to-final mass ratio of the GCs.
}
\label{fig:mass_limit}
\end{figure*}

Using the \citet[][March 2023 update]{baumgardt2021} catalogue, we can construct Figure~\ref{fig:mass_limit}, 
that shows the initial and current GC masses for the entire MW GC population as a function of Galactocentric radius ($R$).
A similar figure was presented in \citet[][see their Fig.~7]{baumgardt2019} but using the semi-major axis instead of $R$.
Figure~\ref{fig:mass_limit} includes also the maximum $z$ extension (i.e., $z_{\rm max}$) as a colour-codded third dimension.
% We adopt an uncertainty of 25\% of the cluster total mass in case current masses were missing in the original catalogue.
We adopt an uncertainty of 25\% of the cluster total mass (i.e., $\sigma_{\rm Current\ Mass} = 0.25 \times M_{\rm Current}$) in case mass uncertainties were missing in the catalogue.

Observationally, clear differences arise using $R$ in the x-axis in this figure. 
First, it is easier to identify the bulge region using $R < 3.5$ kpc, as suggested by \cite{rojasarriagada2020}, and maybe extending up to $\sim 6$ kpc.
In this area, as pointed out by \cite{baumgardt2019}, dynamical processes determine a larger $M_{\rm initial}$ to $M_{\rm current}$ mass differences, 
as low-mass clusters ($\sim 10^4 M_\odot$) are more easily disrupted. 
The halo region also shows a distinct pattern, given the lower mass-loss of clusters and the lower mass observable limit of $\sim 10^3 M_\odot$ hitherto, possibly due to a different origin of these outer GCs.

We can extend this analysis by including the newly discovered GCs and also extend the relationship to the Galactic disk, based on $R$.
By comparing the mean mass loss shift from GCs at the same radius, we derive a rough estimate for the initial mass of Gran~4 to be $\sim 10^6 {\rm M}_\odot$.
Initial masses for the remaining clusters were already derived in the \cite{baumgardt2019} catalogue.
A ``bulge/disk disruption'' region is hatched in grey in Figure~\ref{fig:mass_limit}, highlighting the observational lack of clusters.
Note that this area does not imply that GCs cannot exist in that part of the parameter space, only reveals that observations show no clusters in that area.
A simple linear function $M(R) \sim 10^4 R^{-1} (M_\odot\ {\rm kpc}^{-1})$ was used to separate the parameter space in which GCs are observed. 
This functional form is valid from $R \sim 0.6$ kpc (Gran~1) until the edge of the MW disk at $\sim 18$ kpc \citep{antoja2021}, in which no GC has been observed below this threshold up to date.

%-----------------------------------------------------------------

\section{Discussion and Conclusions}
\label{sec:discussionconclusion}

As a follow-up to our previous work \citep{gran2022}, we have derived a mean radial velocity and metallicity for the cluster Gran~4 and updated the values derived for the Gran~1, 2, 3, and 5. 
Although we have analysed the same MUSE data as in \cite{gran2022}, we have now used the entire MUSE wavelength range rather than only two CaT lines, and we have also upgraded the PMs by using the Gaia DR3 catalogue.

Overall, we find an excellent agreement with previous studies of these clusters, with just a few key differences.
\cite{trincado2022} derived a much lower metallicity for Gran~3 and a different RV for this cluster.
We notice the same trend in the more recent study of \cite{pace2023}.
This issue was corrected when re-analysing the MUSE spectra, due to an incorrect air-to-vacuum wavelength calibration used in \cite{gran2022}, which led to that discrepancy of $\sim20$ km s$^{-1}$ shift between our previous results and the \cite{pace2023}. 
% It also accounts for the metallicity shift, given that the original cluster members were mixed with a different group of field stars.
Considering the metallicity dispersion, our derived RV and metallicity for Gran~3 ($91.57$ km s$^{-1}$ and $-1.63$ dex, respectively, see Table~\ref{table:granes}), 
show remarkable similarity with the ones reported by \cite{trincado2022} ($\sim$$-1.7$ dex and $95.9$ km s$^{-1}$) and \cite{pace2023} ($-1.84$ dex and $90.9$ km s$^{-1}$).
Finally, in the case of Gran~4, we derive a RV and metallicity of [Fe/H]=$-$1.72 dex and RV=$-$265.3 km s$^{-1}$, respectively, 
which are compatible, within errors, with the ones derived in by \citet{pace2023}, i.e., [Fe/H]=$-$1.84 dex and RV=$-$266.4 km s$^{-1}$.
These values agree with the conclusions presented in \cite{pace2023} of Gran~4 being part of the LMS/Wukong merger \citep{conroy2019, naidu2020, malhan2021} or the Helmi stream \citep{helmi1999}. 

% Metalicidad de Gran 1
% [Fe/H] = -1.13 +- 0.05
% RV = 76.98 +- 1.53

% Metalicidad de Gran 2
% [Fe/H] = -1.46 +- 0.08
% RV = 61.24 +- 2.24

% Metalicidad de Gran 3
% [Fe/H] = -1.63 +- 0.07
% RV = 91.57 +- 1.90

% Metalicidad de Gran 4
% [Fe/H] = -1.72 +- 0.09
% RV = -265.28 +- 2.51

% Metalicidad de Gran 5
% [Fe/H] = -1.02 +- 0.05
% RV = -59.19 +- 1.54

Our analysis has proven that Gran~4 is a bonafide GC, previously hidden behind a highly crowded and reddened region towards the Galactic bulge.
An initial mass of $\sim 10^6 {\rm M}_\odot$ was assigned to the cluster based on the $M_{\rm initial}/M_{\rm current}$ ratio of clusters at similar Galactocentric radius.
Furthermore, we present RVs and stellar atmospheric parameters for most of their observed members in a field-of-view of 1 sq arcmin from the centre.
Future planned high-resolution follow-ups will disentangle whether or not the presented GCs host multiple stellar populations.

How many GCs are still missing on the far side of the Galaxy is still uncertain, and locating new GCs is challenging as a result of the high stellar density and reddening towards the Galactic bulge/plane.
Despite our best estimation of $\sim12-26$ GCs still to be discovered, the number heavily relies on critical assumptions like symmetry and cluster survival rate.
However, substantial progress has been achieved in the last few years thanks to the release of the Gaia DR3 catalogue \citep{gaiadr3} and to the improvement in the performances of near-infrared detectors. 

Recently, other stellar tracers have been used as well, shedding light on the spiral structure \citep{sanna2017, minniti2020, jminniti2021}, 
taking advantage of low-extinction windows \citep{saito2020}, or using a variety of distance indicators \citep{hey2023} to map the other side of the Galaxy.
GCs behind the bulge have been discovered and confirmed using joint observations, a number that will only increase with future Gaia DRs.

The promising new generation stellar surveys like WEAVE \citep{jin2023}, 4MOST \citep{dejong2019, lucatello2023}, MOONS \citep{cirasuolo2020, gonzalez2020}, 
in addition to the augmented instrumental and observational capabilities of future facilities, e.g., Vera Rubin \citep{ivezic2019} and Gaia-NIR \citep{hobbs2016}, 
will significantly increase our overall understanding of the MW and its GCs, expanding well-observed clusters towards the plane.

\begin{acknowledgements}
\label{sec:acknowledgements}

We gratefully thank Patrick de Laverny for providing access to the AMBRE synthetic spectra.
FG, GK and VH gratefully acknowledge support from the French National Research Agency (ANR) funded project ``MWDisc'' (ANR-20-CE31-0004) and ``Pristine'' (ANR-18-CE31-0017).
This work was the last part of the Ph.D. thesis of FG, funded by grant CONICYT-PCHA Doctorado Nacional 2017-21171485.
MZ acknowledges support from FONDECYT Regular grant No. 1230731.
ARA acknowledges support from FONDECYT through grant 3180203.
JAC-B acknowledges support from FONDECYT Regular No. 1220083.
EV acknowledges the Excellence Cluster ORIGINS Funded by the Deutsche Forschungsgemeinschaft (DFG, German Research Foundation) under Germany’s Excellence Strategy – EXC-2094-390783311. 
% MDL is funded by ANID, Millenium Science Initiative, ICN12\_009.
Support for MZ, ARA, RC, and MDL is provided by the Ministerio de Ciencia, Tecnología, Conocimiento e Innovación/Agencia Nacional de Investigación y Desarrollo (ANID) Programa Iniciativa Científica Milenio through grant IC120009, awarded to the Millennium Institute of Astrophysics (M.A.S.) and the ANID BASAL Center for Astrophysics and Associated Technologies (CATA) through grant FB210003.\\

We gratefully acknowledge the use of data from the VVV ESO Public Survey program ID 179.B-2002 taken with the VISTA telescope, 
and data products from the Cambridge Astronomical Survey Unit (CASU). 
The VVV Survey data are made public at the ESO Archive. 
Based on observations taken within the ESO VISTA Public Survey VVV, Program ID 179.B-2002.\\

This work has made use of data from the European Space Agency (ESA) mission {\it Gaia} (\url{https://www.cosmos.esa.int/gaia}), 
processed by the {\it Gaia} Data Processing and Analysis Consortium (DPAC,
\url{https://www.cosmos.esa.int/web/gaia/dpac/consortium}). 
Funding for the DPAC has been provided by national institutions, in particular the institutions participating in the {\it Gaia} Multilateral Agreement.\\

This research has made use of the VizieR catalogue access tool, CDS, Strasbourg, France (DOI : 10.26093/cds/vizier). 
The original description of the VizieR service was published in \cite{ochsenbein2000}.
This research made use of: TOPCAT \citep{topcat}, IPython/Jupyter \citep{perez2007, kluyver2016}, numpy \citep{harris2020}, 
matplotlib \citep{hunter2007}, Astropy, a community developed core Python package for Astronomy \citep{astropy1, astropy2, astropy3},
Photutils, an Astropy package for detection and photometry of astronomical sources \citep{bradley2023},
galpy: A Python Library for Galactic Dynamics \citep{bovy2015}, specutils \citep{earl2021}, gala \citep{pricewhelan2017, pricewhelan2020},
and {\tt Uncertainties}: a Python package for calculations with uncertainties (Eric O. LEBIGOT, \url{http://pythonhosted.org/uncertainties/}).
This research has made use of NASA’s Astrophysics Data System.

\end{acknowledgements}

\bibliographystyle{aa} % style aa.bst
\bibliography{biblio.bib} % your references Yourfile.bib

\begin{thebibliography}{119}
\expandafter\ifx\csname natexlab\endcsname\relax\def\natexlab#1{#1}\fi

\bibitem[{{Abell}(1955)}]{pal1}
{Abell}, G.~O. 1955, \pasp, 67, 258

\bibitem[{{Arakelyan} {et~al.}(2018){Arakelyan}, {Pilipenko}, \&
  {Libeskind}}]{arakelyan2018}
{Arakelyan}, N.~R., {Pilipenko}, S.~V., \& {Libeskind}, N.~I. 2018, \mnras,
  481, 918

\bibitem[{{Astropy Collaboration} {et~al.}(2022){Astropy Collaboration},
  {Price-Whelan}, {Lim}, {Earl}, {Starkman}, {Bradley}, {Shupe}, {Patil},
  {Corrales}, {Brasseur}, {N{"o}the}, {Donath}, {Tollerud}, {Morris},
  {Ginsburg}, {Vaher}, {Weaver}, {Tocknell}, {Jamieson}, {van Kerkwijk},
  {Robitaille}, {Merry}, {Bachetti}, {G{"u}nther}, {Aldcroft},
  {Alvarado-Montes}, {Archibald}, {B{'o}di}, {Bapat}, {Barentsen}, {Baz{'a}n},
  {Biswas}, {Boquien}, {Burke}, {Cara}, {Cara}, {Conroy}, {Conseil}, {Craig},
  {Cross}, {Cruz}, {D'Eugenio}, {Dencheva}, {Devillepoix}, {Dietrich},
  {Eigenbrot}, {Erben}, {Ferreira}, {Foreman-Mackey}, {Fox}, {Freij}, {Garg},
  {Geda}, {Glattly}, {Gondhalekar}, {Gordon}, {Grant}, {Greenfield}, {Groener},
  {Guest}, {Gurovich}, {Handberg}, {Hart}, {Hatfield-Dodds}, {Homeier},
  {Hosseinzadeh}, {Jenness}, {Jones}, {Joseph}, {Kalmbach}, {Karamehmetoglu},
  {Ka{l}uszy{'n}ski}, {Kelley}, {Kern}, {Kerzendorf}, {Koch}, {Kulumani},
  {Lee}, {Ly}, {Ma}, {MacBride}, {Maljaars}, {Muna}, {Murphy}, {Norman},
  {O'Steen}, {Oman}, {Pacifici}, {Pascual}, {Pascual-Granado}, {Patil},
  {Perren}, {Pickering}, {Rastogi}, {Roulston}, {Ryan}, {Rykoff}, {Sabater},
  {Sakurikar}, {Salgado}, {Sanghi}, {Saunders}, {Savchenko}, {Schwardt},
  {Seifert-Eckert}, {Shih}, {Jain}, {Shukla}, {Sick}, {Simpson},
  {Singanamalla}, {Singer}, {Singhal}, {Sinha}, {Sip{H{o}}cz}, {Spitler},
  {Stansby}, {Streicher}, {{{S}}umak}, {Swinbank}, {Taranu}, {Tewary},
  {Tremblay}, {Val-Borro}, {Van Kooten}, {Vasovi{'c}}, {Verma}, {de Miranda
  Cardoso}, {Williams}, {Wilson}, {Winkel}, {Wood-Vasey}, {Xue}, {Yoachim},
  {Zhang}, {Zonca}, \& {Astropy Project Contributors}}]{astropy3}
{Astropy Collaboration}, {Price-Whelan}, A.~M., {Lim}, P.~L., {et~al.} 2022,
  apj, 935, 167

\bibitem[{{Astropy Collaboration} {et~al.}(2018){Astropy Collaboration},
  {Price-Whelan}, {Sip{\H{o}}cz}, {G{\"u}nther}, {Lim}, {Crawford}, {Conseil},
  {Shupe}, {Craig}, {Dencheva}, {Ginsburg}, {Vand erPlas}, {Bradley},
  {P{\'e}rez-Su{\'a}rez}, {de Val-Borro}, {Aldcroft}, {Cruz}, {Robitaille},
  {Tollerud}, {Ardelean}, {Babej}, {Bach}, {Bachetti}, {Bakanov}, {Bamford},
  {Barentsen}, {Barmby}, {Baumbach}, {Berry}, {Biscani}, {Boquien}, {Bostroem},
  {Bouma}, {Brammer}, {Bray}, {Breytenbach}, {Buddelmeijer}, {Burke},
  {Calderone}, {Cano Rodr{\'\i}guez}, {Cara}, {Cardoso}, {Cheedella}, {Copin},
  {Corrales}, {Crichton}, {D'Avella}, {Deil}, {Depagne}, {Dietrich}, {Donath},
  {Droettboom}, {Earl}, {Erben}, {Fabbro}, {Ferreira}, {Finethy}, {Fox},
  {Garrison}, {Gibbons}, {Goldstein}, {Gommers}, {Greco}, {Greenfield},
  {Groener}, {Grollier}, {Hagen}, {Hirst}, {Homeier}, {Horton}, {Hosseinzadeh},
  {Hu}, {Hunkeler}, {Ivezi{\'c}}, {Jain}, {Jenness}, {Kanarek}, {Kendrew},
  {Kern}, {Kerzendorf}, {Khvalko}, {King}, {Kirkby}, {Kulkarni}, {Kumar},
  {Lee}, {Lenz}, {Littlefair}, {Ma}, {Macleod}, {Mastropietro}, {McCully},
  {Montagnac}, {Morris}, {Mueller}, {Mumford}, {Muna}, {Murphy}, {Nelson},
  {Nguyen}, {Ninan}, {N{\"o}the}, {Ogaz}, {Oh}, {Parejko}, {Parley}, {Pascual},
  {Patil}, {Patil}, {Plunkett}, {Prochaska}, {Rastogi}, {Reddy Janga},
  {Sabater}, {Sakurikar}, {Seifert}, {Sherbert}, {Sherwood-Taylor}, {Shih},
  {Sick}, {Silbiger}, {Singanamalla}, {Singer}, {Sladen}, {Sooley},
  {Sornarajah}, {Streicher}, {Teuben}, {Thomas}, {Tremblay}, {Turner},
  {Terr{\'o}n}, {van Kerkwijk}, {de la Vega}, {Watkins}, {Weaver}, {Whitmore},
  {Woillez}, {Zabalza}, \& {Astropy Contributors}}]{astropy2}
{Astropy Collaboration}, {Price-Whelan}, A.~M., {Sip{\H{o}}cz}, B.~M., {et~al.}
  2018, \aj, 156, 123

\bibitem[{{Astropy Collaboration} {et~al.}(2013){Astropy Collaboration},
  {Robitaille}, {Tollerud}, {Greenfield}, {Droettboom}, {Bray}, {Aldcroft},
  {Davis}, {Ginsburg}, {Price-Whelan}, {Kerzendorf}, {Conley}, {Crighton},
  {Barbary}, {Muna}, {Ferguson}, {Grollier}, {Parikh}, {Nair}, {Unther},
  {Deil}, {Woillez}, {Conseil}, {Kramer}, {Turner}, {Singer}, {Fox}, {Weaver},
  {Zabalza}, {Edwards}, {Azalee Bostroem}, {Burke}, {Casey}, {Crawford},
  {Dencheva}, {Ely}, {Jenness}, {Labrie}, {Lim}, {Pierfederici}, {Pontzen},
  {Ptak}, {Refsdal}, {Servillat}, \& {Streicher}}]{astropy1}
{Astropy Collaboration}, {Robitaille}, T.~P., {Tollerud}, E.~J., {et~al.} 2013,
  \aap, 558, A33

\bibitem[{{Bacon} {et~al.}(2010){Bacon}, {Accardo}, {Adjali}, {Anwand},
  {Bauer}, {Biswas}, {Blaizot}, {Boudon}, {Brau-Nogue}, {Brinchmann},
  {Caillier}, {Capoani}, {Carollo}, {Contini}, {Couderc}, {Daguis{\'e}},
  {Deiries}, {Delabre}, {Dreizler}, {Dubois}, {Dupieux}, {Dupuy}, {Emsellem},
  {Fechner}, {Fleischmann}, {Fran{\c{c}}ois}, {Gallou}, {Gharsa}, {Glindemann},
  {Gojak}, {Guiderdoni}, {Hansali}, {Hahn}, {Jarno}, {Kelz}, {Koehler},
  {Kosmalski}, {Laurent}, {Le Floch}, {Lilly}, {Lizon}, {Loupias}, {Manescau},
  {Monstein}, {Nicklas}, {Olaya}, {Pares}, {Pasquini}, {P{\'e}contal-Rousset},
  {Pell{\'o}}, {Petit}, {Popow}, {Reiss}, {Remillieux}, {Renault}, {Roth},
  {Rupprecht}, {Serre}, {Schaye}, {Soucail}, {Steinmetz}, {Streicher}, {Stuik},
  {Valentin}, {Vernet}, {Weilbacher}, {Wisotzki}, \& {Yerle}}]{bacon2010}
{Bacon}, R., {Accardo}, M., {Adjali}, L., {et~al.} 2010, in Society of
  Photo-Optical Instrumentation Engineers (SPIE) Conference Series, Vol. 7735,
  Ground-based and Airborne Instrumentation for Astronomy III, ed. I.~S.
  {McLean}, S.~K. {Ramsay}, \& H.~{Takami}, 773508

\bibitem[{{Bailin}(2019)}]{bailin2019}
{Bailin}, J. 2019, \apjs, 245, 5

\bibitem[{{Bailin} \& {von Klar}(2022)}]{bailin2022}
{Bailin}, J. \& {von Klar}, R. 2022, \apj, 925, 36

\bibitem[{{Bastian} \& {Lardo}(2018)}]{bastian2018}
{Bastian}, N. \& {Lardo}, C. 2018, \araa, 56, 83

\bibitem[{{Baumgardt} {et~al.}(2019{\natexlab{a}}){Baumgardt}, {Hilker},
  {Sollima}, \& {Bellini}}]{baumgardt2018}
{Baumgardt}, H., {Hilker}, M., {Sollima}, A., \& {Bellini}, A.
  2019{\natexlab{a}}, \mnras, 482, 5138

\bibitem[{{Baumgardt} {et~al.}(2019{\natexlab{b}}){Baumgardt}, {Hilker},
  {Sollima}, \& {Bellini}}]{baumgardt2019}
{Baumgardt}, H., {Hilker}, M., {Sollima}, A., \& {Bellini}, A.
  2019{\natexlab{b}}, \mnras, 482, 5138

\bibitem[{{Baumgardt} \& {Makino}(2003)}]{baumgardt2003}
{Baumgardt}, H. \& {Makino}, J. 2003, \mnras, 340, 227

\bibitem[{{Baumgardt} \& {Vasiliev}(2021)}]{baumgardt2021}
{Baumgardt}, H. \& {Vasiliev}, E. 2021, \mnras, 505, 5957

\bibitem[{{Bovy}(2015)}]{bovy2015}
{Bovy}, J. 2015, \apjs, 216, 29

\bibitem[{Bradley {et~al.}(2023)Bradley, Sip{\H o}cz, Robitaille, Tollerud,
  Vin{\'{\i}}cius, Deil, Barbary, Wilson, Busko, Donath, G{\"u}nther, Cara,
  Lim, Me{\ss}linger, Conseil, Bostroem, Droettboom, Bray, Bratholm, Barentsen,
  Craig, Rathi, Pascual, Perren, Georgiev, de~Val-Borro, Kerzendorf, Bach,
  Quint, \& Souchereau}]{bradley2023}
Bradley, L., Sip{\H o}cz, B., Robitaille, T., {et~al.} 2023, astropy/photutils:
  1.8.0

\bibitem[{Bradley {et~al.}(2022)Bradley, Sipőcz, Robitaille, Tollerud,
  Vinícius, Deil, Barbary, Wilson, Busko, Donath, Günther, Cara, Lim,
  Meßlinger, Conseil, Bostroem, Droettboom, Bray, Bratholm, Barentsen, Craig,
  Rathi, Pascual, Perren, Georgiev, de~Val-Borro, Kerzendorf, Bach, Quint, \&
  Souchereau}]{photutils}
Bradley, L., Sipőcz, B., Robitaille, T., {et~al.} 2022, astropy/photutils:
  1.5.0

\bibitem[{{Bragaglia}(2018)}]{bragaglia2018}
{Bragaglia}, A. 2018, in Astrometry and Astrophysics in the Gaia Sky, ed.
  A.~{Recio-Blanco}, P.~{de Laverny}, A.~G.~A. {Brown}, \& T.~{Prusti}, Vol.
  330, 119--126

\bibitem[{{Bressan} {et~al.}(2012){Bressan}, {Marigo}, {Girardi}, {Salasnich},
  {Dal Cero}, {Rubele}, \& {Nanni}}]{bressan2012}
{Bressan}, A., {Marigo}, P., {Girardi}, L., {et~al.} 2012, \mnras, 427, 127

\bibitem[{{Brodie} \& {Strader}(2006)}]{brodie2006}
{Brodie}, J.~P. \& {Strader}, J. 2006, \araa, 44, 193

\bibitem[{{Callingham} {et~al.}(2022){Callingham}, {Cautun}, {Deason}, {Frenk},
  {Grand}, \& {Marinacci}}]{callingham2022}
{Callingham}, T.~M., {Cautun}, M., {Deason}, A.~J., {et~al.} 2022, \mnras, 513,
  4107

\bibitem[{{Cantat-Gaudin}(2022)}]{cantat-gaudin2022}
{Cantat-Gaudin}, T. 2022, Universe, 8, 111

\bibitem[{{Cantat-Gaudin} \& {Anders}(2020)}]{cantat-gaudin2020a}
{Cantat-Gaudin}, T. \& {Anders}, F. 2020, \aap, 633, A99

\bibitem[{{Cantat-Gaudin} {et~al.}(2020){Cantat-Gaudin}, {Anders},
  {Castro-Ginard}, {Jordi}, {Romero-G{\'o}mez}, {Soubiran}, {Casamiquela},
  {Tarricq}, {Moitinho}, {Vallenari}, {Bragaglia}, {Krone-Martins}, \&
  {Kounkel}}]{cantat-gaudin2020b}
{Cantat-Gaudin}, T., {Anders}, F., {Castro-Ginard}, A., {et~al.} 2020, \aap,
  640, A1

\bibitem[{{Cantat-Gaudin} {et~al.}(2019){Cantat-Gaudin}, {Krone-Martins},
  {Sedaghat}, {Farahi}, {de Souza}, {Skalidis}, {Malz}, {Mac{\^e}do}, {Moews},
  {Jordi}, {Moitinho}, {Castro-Ginard}, {Ishida}, {Heneka}, {Boucaud}, \&
  {Trindade}}]{cantat-gaudin2019}
{Cantat-Gaudin}, T., {Krone-Martins}, A., {Sedaghat}, N., {et~al.} 2019, \aap,
  624, A126

\bibitem[{{Carlberg} \& {Keating}(2022)}]{carlberg2022}
{Carlberg}, R.~G. \& {Keating}, L.~C. 2022, \apj, 924, 77

\bibitem[{{Carraro}(2005)}]{whiting1}
{Carraro}, G. 2005, \apjl, 621, L61

\bibitem[{{Castro-Ginard} {et~al.}(2020){Castro-Ginard}, {Jordi}, {Luri},
  {{\'A}lvarez Cid-Fuentes}, {Casamiquela}, {Anders}, {Cantat-Gaudin},
  {Mongui{\'o}}, {Balaguer-N{\'u}{\~n}ez}, {Sol{\`a}}, \&
  {Badia}}]{castro-ginard2020}
{Castro-Ginard}, A., {Jordi}, C., {Luri}, X., {et~al.} 2020, \aap, 635, A45

\bibitem[{{Castro-Ginard} {et~al.}(2022){Castro-Ginard}, {Jordi}, {Luri},
  {Cantat-Gaudin}, {Carrasco}, {Casamiquela}, {Anders},
  {Balaguer-N{\'u}{\~n}ez}, \& {Badia}}]{castro-ginard2022}
{Castro-Ginard}, A., {Jordi}, C., {Luri}, X., {et~al.} 2022, \aap, 661, A118

\bibitem[{{Castro-Ginard} {et~al.}(2018){Castro-Ginard}, {Jordi}, {Luri},
  {Julbe}, {Morvan}, {Balaguer-N{\'u}{\~n}ez}, \&
  {Cantat-Gaudin}}]{castro-ginard2018}
{Castro-Ginard}, A., {Jordi}, C., {Luri}, X., {et~al.} 2018, \aap, 618, A59

\bibitem[{{Choi} {et~al.}(2018){Choi}, {Conroy}, {Ting}, {Cargile}, {Dotter},
  \& {Johnson}}]{choi2018}
{Choi}, J., {Conroy}, C., {Ting}, Y.-S., {et~al.} 2018, \apj, 863, 65

\bibitem[{{Cirasuolo} {et~al.}(2020){Cirasuolo}, {Fairley}, {Rees}, {Gonzalez},
  {Taylor}, {Maiolino}, {Afonso}, {Evans}, {Flores}, {Lilly}, \&
  et~al.}]{cirasuolo2020}
{Cirasuolo}, M., {Fairley}, A., {Rees}, P., {et~al.} 2020, The Messenger, 180,
  10

\bibitem[{{Conroy} {et~al.}(2019){Conroy}, {Bonaca}, {Cargile}, {Johnson},
  {Caldwell}, {Naidu}, {Zaritsky}, {Fabricant}, {Moran}, {Rhee},
  {Szentgyorgyi}, {Berlind}, {Calkins}, {Kattner}, \& {Ly}}]{conroy2019}
{Conroy}, C., {Bonaca}, A., {Cargile}, P., {et~al.} 2019, \apj, 883, 107

\bibitem[{{Contenta} {et~al.}(2017){Contenta}, {Gieles}, {Balbinot}, \&
  {Collins}}]{contenta2017}
{Contenta}, F., {Gieles}, M., {Balbinot}, E., \& {Collins}, M. L.~M. 2017,
  \mnras, 466, 1741

\bibitem[{{Contreras Ramos} {et~al.}(2017){Contreras Ramos}, {Zoccali},
  {Rojas}, {Rojas-Arriagada}, {G{\'a}rate}, {Huijse}, {Gran}, {Soto},
  {Valcarce}, {Est{\'e}vez}, \& {Minniti}}]{contrerasramos2017}
{Contreras Ramos}, R., {Zoccali}, M., {Rojas}, F., {et~al.} 2017, \aap, 608,
  A140

\bibitem[{{de Jong} {et~al.}(2019){de Jong}, {Agertz}, {Berbel}, {Aird},
  {Alexander}, {Amarsi}, {Anders}, {Andrae}, {Ansarinejad}, {Ansorge}, \&
  et~al.}]{dejong2019}
{de Jong}, R.~S., {Agertz}, O., {Berbel}, A.~A., {et~al.} 2019, The Messenger,
  175, 3

\bibitem[{{de Laverny} {et~al.}(2013){de Laverny}, {Recio-Blanco}, {Worley},
  {De Pascale}, {Hill}, \& {Bijaoui}}]{delaverny2013}
{de Laverny}, P., {Recio-Blanco}, A., {Worley}, C.~C., {et~al.} 2013, The
  Messenger, 153, 18

\bibitem[{{de Laverny} {et~al.}(2012){de Laverny}, {Recio-Blanco}, {Worley}, \&
  {Plez}}]{ambre}
{de Laverny}, P., {Recio-Blanco}, A., {Worley}, C.~C., \& {Plez}, B. 2012,
  \aap, 544, A126

\bibitem[{Earl {et~al.}(2021)Earl, Tollerud, Jones, Kerzendorf, shaileshahuja,
  Busko, D'Avella, O'Steen, Robitaille, Ginsburg, Homeier, Sipőcz, Averbukh,
  Ogaz, Geda, Tocknell, Davies, Günther, Cherinka, Barbary, Foster,
  Droettboom, Torres, Bray, Casey, Teuben, Crawford, Ferguson, Cruz, \&
  Ogle}]{earl2021}
Earl, N., Tollerud, E., Jones, C., {et~al.} 2021, astropy/specutils: v1.2

\bibitem[{{Fern{\'a}ndez-Trincado} {et~al.}(2022){Fern{\'a}ndez-Trincado},
  {Minniti}, {Garro}, \& {Villanova}}]{trincado2022}
{Fern{\'a}ndez-Trincado}, J.~G., {Minniti}, D., {Garro}, E.~R., \& {Villanova},
  S. 2022, \aap, 657, A84

\bibitem[{{Ferrone} {et~al.}(2023){Ferrone}, {Di Matteo},
  {Mastrobuono-Battisti}, {Haywood}, {Snaith}, {Montuori}, {Khoperskov}, \&
  {Valls-Gabaud}}]{ferrone2023}
{Ferrone}, S., {Di Matteo}, P., {Mastrobuono-Battisti}, A., {et~al.} 2023,
  \aap, 673, A44

\bibitem[{{Forbes} {et~al.}(2018){Forbes}, {Bastian}, {Gieles}, {Crain},
  {Kruijssen}, {Larsen}, {Ploeckinger}, {Agertz}, {Trenti}, {Ferguson},
  {Pfeffer}, \& {Gnedin}}]{forbes2018}
{Forbes}, D.~A., {Bastian}, N., {Gieles}, M., {et~al.} 2018, Proceedings of the
  Royal Society of London Series A, 474, 20170616

\bibitem[{{Forbes} \& {Bridges}(2010)}]{forbes2010}
{Forbes}, D.~A. \& {Bridges}, T. 2010, \mnras, 404, 1203

\bibitem[{{Foreman-Mackey} {et~al.}(2013){Foreman-Mackey}, {Hogg}, {Lang}, \&
  {Goodman}}]{emcee}
{Foreman-Mackey}, D., {Hogg}, D.~W., {Lang}, D., \& {Goodman}, J. 2013, PASP,
  125, 306

\bibitem[{{Gaia Collaboration} {et~al.}(2021){Gaia Collaboration}, {Antoja},
  {McMillan}, {Kordopatis}, {Ramos}, {Helmi}, {Balbinot}, {Cantat-Gaudin},
  {Chemin}, {Figueras}, \& et~al.}]{antoja2021}
{Gaia Collaboration}, {Antoja}, T., {McMillan}, P.~J., {et~al.} 2021, \aap,
  649, A8

\bibitem[{{Gaia Collaboration} {et~al.}(2018){Gaia Collaboration}, {Brown},
  {Vallenari}, {Prusti}, {de Bruijne}, {Babusiaux}, {Bailer-Jones}, {Biermann},
  {Evans}, {Eyer}, \& et~al.}]{gaiadr2}
{Gaia Collaboration}, {Brown}, A.~G.~A., {Vallenari}, A., {et~al.} 2018, \aap,
  616, A1

\bibitem[{{Gaia Collaboration} {et~al.}(2016){Gaia Collaboration}, {Prusti},
  {de Bruijne}, {Brown}, {Vallenari}, {Babusiaux}, {Bailer-Jones}, {Bastian},
  {Biermann}, {Evans}, \& et~al.}]{gaia}
{Gaia Collaboration}, {Prusti}, T., {de Bruijne}, J.~H.~J., {et~al.} 2016,
  \aap, 595, A1

\bibitem[{{Gaia Collaboration} {et~al.}(2022){Gaia Collaboration}, {Vallenari},
  {Brown}, {Prusti}, {de Bruijne}, {Arenou}, {Babusiaux}, {Biermann},
  {Creevey}, {Ducourant}, \& et~al.}]{gaiadr3}
{Gaia Collaboration}, {Vallenari}, A., {Brown}, A.~G.~A., {et~al.} 2022, arXiv
  e-prints, arXiv:2208.00211

\bibitem[{{Garro} {et~al.}(2022{\natexlab{a}}){Garro}, {Minniti}, {Alessi},
  {Patchick}, {Kronberger}, {Alonso-Garc{\'\i}a}, {Fern{\'a}ndez-Trincado},
  {G{\'o}mez}, {Hempel}, {Pullen}, {Saito}, {Ripepi}, \& {Zelada
  Bacigalupo}}]{garro2022a}
{Garro}, E.~R., {Minniti}, D., {Alessi}, B., {et~al.} 2022{\natexlab{a}}, \aap,
  659, A155

\bibitem[{{Garro} {et~al.}(2020){Garro}, {Minniti}, {G{\'o}mez},
  {Alonso-Garc{\'\i}a}, {Barb{\'a}}, {Barbuy}, {Clari{\'a}}, {Chen{\'e}},
  {Dias}, {Hempel}, {Ivanov}, {Lucas}, {Majaess}, {Mauro}, {Moni Bidin},
  {Palma}, {Pullen}, {Saito}, {Smith}, {Surot}, {Ram{\'\i}rez Alegr{\'\i}a},
  {Rejkuba}, {Ripepi}, \& {Fern{\'a}ndez Trincado}}]{garro2020}
{Garro}, E.~R., {Minniti}, D., {G{\'o}mez}, M., {et~al.} 2020, \aap, 642, L19

\bibitem[{{Garro} {et~al.}(2022{\natexlab{b}}){Garro}, {Minniti}, {G{\'o}mez},
  {Fern{\'a}ndez-Trincado}, {Alonso-Garc{\'\i}a}, {Hempel}, \& {Zelada
  Bacigalupo}}]{garro2022b}
{Garro}, E.~R., {Minniti}, D., {G{\'o}mez}, M., {et~al.} 2022{\natexlab{b}},
  \aap, 662, A95

\bibitem[{{Gonzalez} {et~al.}(2020){Gonzalez}, {Mucciarelli}, {Origlia},
  {Schultheis}, {Caffau}, {Di Matteo}, {Randich}, {Recio-Blanco}, {Zoccali},
  {Bonifacio}, {Dalessandro}, {Schiavon}, {Pancino}, {Taylor}, {Valenti},
  {Rojas-Arriagada}, {Sacco}, {Biazzo}, {Bellazzini}, {Cioni}, {Clementini},
  {Contreras Ramos}, {de Laverny}, {Evans}, {Haywood}, {Hill}, {Ibata},
  {Lucatello}, {Magrini}, {Martin}, {Nisini}, {Sanna}, {Cirasuolo}, {Maiolino},
  {Afonso}, {Lilly}, {Flores}, {Oliva}, {Paltani}, \& {Vanzi}}]{gonzalez2020}
{Gonzalez}, O.~A., {Mucciarelli}, A., {Origlia}, L., {et~al.} 2020, The
  Messenger, 180, 18

\bibitem[{{Gran} {et~al.}(2019){Gran}, {Zoccali}, {Contreras Ramos}, {Valenti},
  {Rojas-Arriagada}, {Carballo-Bello}, {Alonso-Garcia}, {Minniti}, {Rejkuba},
  \& {Surot}}]{gran2019}
{Gran}, F., {Zoccali}, M., {Contreras Ramos}, R., {et~al.} 2019, \aap, 628, A45

\bibitem[{{Gran} {et~al.}(2022){Gran}, {Zoccali}, {Saviane}, {Valenti},
  {Rojas-Arriagada}, {Contreras Ramos}, {Hartke}, {Carballo-Bello},
  {Navarrete}, {Rejkuba}, \& {Olivares Carvajal}}]{gran2022}
{Gran}, F., {Zoccali}, M., {Saviane}, I., {et~al.} 2022, \mnras, 509, 4962

\bibitem[{{Gratton} {et~al.}(2019){Gratton}, {Bragaglia}, {Carretta},
  {D'Orazi}, {Lucatello}, \& {Sollima}}]{gratton2019}
{Gratton}, R., {Bragaglia}, A., {Carretta}, E., {et~al.} 2019, \aapr, 27, 8

\bibitem[{{Gratton} {et~al.}(2004){Gratton}, {Sneden}, \&
  {Carretta}}]{gratton2004}
{Gratton}, R., {Sneden}, C., \& {Carretta}, E. 2004, \araa, 42, 385

\bibitem[{{Hammer} {et~al.}(2023){Hammer}, {Li}, {Mamon}, {Pawlowski},
  {Bonifacio}, {Jiao}, {Wang}, {Wang}, \& {Yang}}]{hammer2023}
{Hammer}, F., {Li}, H., {Mamon}, G.~A., {et~al.} 2023, \mnras, 519, 5059

\bibitem[{Harris {et~al.}(2020)Harris, Millman, van~der Walt, Gommers,
  Virtanen, Cournapeau, Wieser, Taylor, Berg, Smith, Kern, Picus, Hoyer, van
  Kerkwijk, Brett, Haldane, del R{\'{i}}o, Wiebe, Peterson,
  G{\'{e}}rard-Marchant, Sheppard, Reddy, Weckesser, Abbasi, Gohlke, \&
  Oliphant}]{harris2020}
Harris, C.~R., Millman, K.~J., van~der Walt, S.~J., {et~al.} 2020, Nature, 585,
  357

\bibitem[{{Harris} \& {Racine}(1979)}]{harris1979}
{Harris}, W.~E. \& {Racine}, R. 1979, \araa, 17, 241

\bibitem[{{Hartke} {et~al.}(2020){Hartke}, {Kakkad}, {Reyes}, {Moya-Sierralta},
  {Reyes}, {Kravtsov}, {Kolb}, \& {Selman}}]{hartke2020}
{Hartke}, J., {Kakkad}, D., {Reyes}, C., {et~al.} 2020, in Society of
  Photo-Optical Instrumentation Engineers (SPIE) Conference Series, Vol. 11448,
  Society of Photo-Optical Instrumentation Engineers (SPIE) Conference Series,
  114480V

\bibitem[{{He} {et~al.}(2023){He}, {Liu}, {Luo}, {Wang}, \& {Jiang}}]{he2023}
{He}, Z., {Liu}, X., {Luo}, Y., {Wang}, K., \& {Jiang}, Q. 2023, \apjs, 264, 8

\bibitem[{{He} {et~al.}(2022){He}, {Wang}, {Luo}, {Li}, {Liu}, \&
  {Jiang}}]{he2022}
{He}, Z., {Wang}, K., {Luo}, Y., {et~al.} 2022, \apjs, 262, 7

\bibitem[{{He} {et~al.}(2021){He}, {Xu}, {Hao}, {Wu}, \& {Li}}]{he2021}
{He}, Z.-H., {Xu}, Y., {Hao}, C.-J., {Wu}, Z.-Y., \& {Li}, J.-J. 2021, Research
  in Astronomy and Astrophysics, 21, 093

\bibitem[{{Helmi} {et~al.}(1999){Helmi}, {White}, {de Zeeuw}, \&
  {Zhao}}]{helmi1999}
{Helmi}, A., {White}, S. D.~M., {de Zeeuw}, P.~T., \& {Zhao}, H. 1999, \nat,
  402, 53

\bibitem[{{Hey} {et~al.}(2023){Hey}, {Huber}, {Shappee}, {Bland-Hawthorn},
  {Tepper-Garc{\'\i}a}, {Sanderson}, {Chakrabarti}, {Saunders}, {Hunt},
  {Bedding}, \& {Tonry}}]{hey2023}
{Hey}, D.~R., {Huber}, D., {Shappee}, B.~J., {et~al.} 2023, arXiv e-prints,
  arXiv:2305.19319

\bibitem[{{Hobbs} {et~al.}(2016){Hobbs}, {H{\o}g}, {Mora}, {Crowley},
  {McMillan}, {Ranalli}, {Heiter}, {Jordi}, {Hambly}, {Church}, {Anthony},
  {Tanga}, {Chemin}, {Portell}, {Jim{\'e}nez-Esteban}, {Klioner}, {Mignard},
  {Fynbo}, {Wyrzykowski}, {Rybicki}, {Anderson}, {Cellino}, {Fabricius},
  {Davidson}, \& {Lindegren}}]{hobbs2016}
{Hobbs}, D., {H{\o}g}, E., {Mora}, A., {et~al.} 2016, arXiv e-prints,
  arXiv:1609.07325

\bibitem[{{Huang} \& {Koposov}(2021)}]{huang2021}
{Huang}, K.-W. \& {Koposov}, S.~E. 2021, \mnras, 500, 986

\bibitem[{{Hughes} {et~al.}(2020){Hughes}, {Pfeffer}, {Martig}, {Reina-Campos},
  {Bastian}, {Crain}, \& {Kruijssen}}]{hughes2020}
{Hughes}, M.~E., {Pfeffer}, J.~L., {Martig}, M., {et~al.} 2020, \mnras, 491,
  4012

\bibitem[{{Hunt} \& {Reffert}(2021)}]{hunt2021}
{Hunt}, E.~L. \& {Reffert}, S. 2021, \aap, 646, A104

\bibitem[{{Hunt} \& {Reffert}(2023)}]{hunt2023}
{Hunt}, E.~L. \& {Reffert}, S. 2023, \aap, 673, A114

\bibitem[{Hunter(2007)}]{hunter2007}
Hunter, J.~D. 2007, Computing in Science \& Engineering, 9, 90

\bibitem[{{Husser} {et~al.}(2020){Husser}, {Latour}, {Brinchmann}, {Dreizler},
  {Giesers}, {G{\"o}ttgens}, {Kamann}, {Roth}, {Weilbacher}, \&
  {Wendt}}]{husser2020}
{Husser}, T.-O., {Latour}, M., {Brinchmann}, J., {et~al.} 2020, \aap, 635, A114

\bibitem[{{Ishchenko} {et~al.}(2023){Ishchenko}, {Sobolenko}, {Berczik},
  {Khoperskov}, {Omarov}, {Sobodar}, \& {Makukov}}]{ishchenko2023}
{Ishchenko}, M., {Sobolenko}, M., {Berczik}, P., {et~al.} 2023, arXiv e-prints,
  arXiv:2304.03547

\bibitem[{{Ivezi{\'c}} {et~al.}(2019){Ivezi{\'c}}, {Kahn}, {Tyson}, {Abel},
  {Acosta}, {Allsman}, {Alonso}, {AlSayyad}, {Anderson}, {Andrew}, \&
  et~al.}]{ivezic2019}
{Ivezi{\'c}}, {\v{Z}}., {Kahn}, S.~M., {Tyson}, J.~A., {et~al.} 2019, \apj,
  873, 111

\bibitem[{{Jin} {et~al.}(2023){Jin}, {Trager}, {Dalton}, {Aguerri}, {Drew},
  {Falc{\'o}n-Barroso}, {G{\"a}nsicke}, {Hill}, {Iovino}, {Pieri}, \&
  et~al.}]{jin2023}
{Jin}, S., {Trager}, S.~C., {Dalton}, G.~B., {et~al.} 2023, \mnras
  [\eprint[arXiv]{2212.03981}]

\bibitem[{Kluyver {et~al.}(2016)Kluyver, Ragan-Kelley, P{\'e}rez, Granger,
  Bussonnier, Frederic, Kelley, Hamrick, Grout, Corlay, Ivanov, Avila, Abdalla,
  \& Willing}]{kluyver2016}
Kluyver, T., Ragan-Kelley, B., P{\'e}rez, F., {et~al.} 2016, in Positioning and
  Power in Academic Publishing: Players, Agents and Agendas, ed. F.~Loizides \&
  B.~Schmidt, IOS Press, 87 -- 90

\bibitem[{{Koposov} {et~al.}(2017){Koposov}, {Belokurov}, \&
  {Torrealba}}]{koposov2017}
{Koposov}, S.~E., {Belokurov}, V., \& {Torrealba}, G. 2017, \mnras, 470, 2702

\bibitem[{{Kruijssen} {et~al.}(2012){Kruijssen}, {Maschberger}, {Moeckel},
  {Clarke}, {Bastian}, \& {Bonnell}}]{kruijssen2012}
{Kruijssen}, J.~M.~D., {Maschberger}, T., {Moeckel}, N., {et~al.} 2012, \mnras,
  419, 841

\bibitem[{{Kruijssen} {et~al.}(2019{\natexlab{a}}){Kruijssen}, {Pfeffer},
  {Crain}, \& {Bastian}}]{kruijssen2019a}
{Kruijssen}, J.~M.~D., {Pfeffer}, J.~L., {Crain}, R.~A., \& {Bastian}, N.
  2019{\natexlab{a}}, \mnras, 486, 3134

\bibitem[{{Kruijssen} {et~al.}(2019{\natexlab{b}}){Kruijssen}, {Pfeffer},
  {Reina-Campos}, {Crain}, \& {Bastian}}]{kruijssen2019b}
{Kruijssen}, J.~M.~D., {Pfeffer}, J.~L., {Reina-Campos}, M., {Crain}, R.~A., \&
  {Bastian}, N. 2019{\natexlab{b}}, \mnras, 486, 3180

\bibitem[{{Leaman} {et~al.}(2013){Leaman}, {VandenBerg}, \&
  {Mendel}}]{leaman2013}
{Leaman}, R., {VandenBerg}, D.~A., \& {Mendel}, J.~T. 2013, \mnras, 436, 122

\bibitem[{{Lucatello} {et~al.}(2023){Lucatello}, {Bragaglia}, {Vallenari},
  {Cantat-Gaudin}, {Kuzma}, {Guarcello}, {Spina}, {Aguado}, {Carrera},
  {Castro-Ginard}, {Damiani}, {D'Orazi}, {Prisinzano}, {Valenti}, {Alfaro},
  {Balaguer-Nu{\~n}ez}, {Balbinot}, {Barrado}, {Baumgardt}, {Bellazzini},
  {Bonito}, {Bossini}, {Carraro}, {Carretta}, {Catanzaro}, {Casamiquela},
  {Cassisi}, {Dalessandro}, {De Silva}, {Ferguson}, {Ferraro}, {Frasca},
  {Galli}, {Gieles}, {Gran}, {Gratton}, {Hilker}, {Jeffries}, {Jordi}, {Korn},
  {Lanzoni}, {Larsen}, {Lattanzio}, {Lugaro}, {Mapelli}, {Massari}, {Miglio},
  {Miret-Roig}, {Momany}, {Mucciarelli}, {Olivares}, {Pasquato},
  {Roccatagliata}, {Salaris}, {Schiavon}, {Smiljanic}, {Sollima},
  {Tautvai{\v{s}}ien{\.{e}}}, {Varri}, \& {Wright}}]{lucatello2023}
{Lucatello}, S., {Bragaglia}, A., {Vallenari}, A., {et~al.} 2023, The
  Messenger, 190, 13

\bibitem[{{Madore} \& {Arp}(1982)}]{am4}
{Madore}, B.~F. \& {Arp}, H.~C. 1982, \pasp, 94, 40

\bibitem[{{Malhan} {et~al.}(2021){Malhan}, {Yuan}, {Ibata}, {Arentsen},
  {Bellazzini}, \& {Martin}}]{malhan2021}
{Malhan}, K., {Yuan}, Z., {Ibata}, R.~A., {et~al.} 2021, \apj, 920, 51

\bibitem[{{Marigo} {et~al.}(2013){Marigo}, {Bressan}, {Nanni}, {Girardi}, \&
  {Pumo}}]{marigo2013}
{Marigo}, P., {Bressan}, A., {Nanni}, A., {Girardi}, L., \& {Pumo}, M.~L. 2013,
  \mnras, 434, 488

\bibitem[{{Mar{\'\i}n-Franch} {et~al.}(2009){Mar{\'\i}n-Franch}, {Aparicio},
  {Piotto}, {Rosenberg}, {Chaboyer}, {Sarajedini}, {Siegel}, {Anderson},
  {Bedin}, {Dotter}, {Hempel}, {King}, {Majewski}, {Milone}, {Paust}, \&
  {Reid}}]{martinfranch2009}
{Mar{\'\i}n-Franch}, A., {Aparicio}, A., {Piotto}, G., {et~al.} 2009, \apj,
  694, 1498

\bibitem[{{Massari} {et~al.}(2019){Massari}, {Koppelman}, \&
  {Helmi}}]{massari2019}
{Massari}, D., {Koppelman}, H.~H., \& {Helmi}, A. 2019, \aap, 630, L4

\bibitem[{{Minniti} {et~al.}(2021{\natexlab{a}}){Minniti},
  {Fern{\'a}ndez-Trincado}, {Smith}, {Lucas}, {G{\'o}mez}, \&
  {Pullen}}]{minniti2021}
{Minniti}, D., {Fern{\'a}ndez-Trincado}, J.~G., {Smith}, L.~C., {et~al.}
  2021{\natexlab{a}}, \aap, 648, A86

\bibitem[{{Minniti} {et~al.}(2017){Minniti}, {Geisler}, {Alonso-Garc{\'\i}a},
  {Palma}, {Beam{\'\i}n}, {Borissova}, {Catelan}, {Clari{\'a}}, {Cohen},
  {Contreras Ramos}, {Dias}, {Fern{\'a}ndez-Trincado}, {G{\'o}mez}, {Hempel},
  {Ivanov}, {Kurtev}, {Lucas}, {Moni-Bidin}, {Pullen}, {Ram{\'\i}rez
  Alegr{\'\i}a}, {Saito}, \& {Valenti}}]{minniti2017}
{Minniti}, D., {Geisler}, D., {Alonso-Garc{\'\i}a}, J., {et~al.} 2017, \apjl,
  849, L24

\bibitem[{{Minniti} {et~al.}(2010){Minniti}, {Lucas}, {Emerson}, {Saito},
  {Hempel}, {Pietrukowicz}, {Ahumada}, {Alonso}, {Alonso-Garcia}, {Arias},
  {Bandyopadhyay}, {Barb{\'a}}, {Barbuy}, {Bedin}, {Bica}, {Borissova},
  {Bronfman}, {Carraro}, {Catelan}, {Clari{\'a}}, {Cross}, {de Grijs},
  {D{\'e}k{\'a}ny}, {Drew}, {Fari{\~n}a}, {Feinstein}, {Fern{\'a}ndez
  Laj{\'u}s}, {Gamen}, {Geisler}, {Gieren}, {Goldman}, {Gonzalez}, {Gunthardt},
  {Gurovich}, {Hambly}, {Irwin}, {Ivanov}, {Jord{\'a}n}, {Kerins}, {Kinemuchi},
  {Kurtev}, {L{\'o}pez-Corredoira}, {Maccarone}, {Masetti}, {Merlo},
  {Messineo}, {Mirabel}, {Monaco}, {Morelli}, {Padilla}, {Palma}, {Parisi},
  {Pignata}, {Rejkuba}, {Roman-Lopes}, {Sale}, {Schreiber}, {Schr{\"o}der},
  {Smith}, {}, {Soto}, {Tamura}, {Tappert}, {Thompson}, {Toledo}, {Zoccali}, \&
  {Pietrzynski}}]{vvv}
{Minniti}, D., {Lucas}, P.~W., {Emerson}, J.~P., {et~al.} 2010, \na, 15, 433

\bibitem[{{Minniti} {et~al.}(2020){Minniti}, {Sbordone}, {Rojas-Arriagada},
  {Zoccali}, {Contreras Ramos}, {Minniti}, {Marconi}, {Braga}, {Catelan},
  {Duffau}, {Gieren}, \& {Valcarce}}]{minniti2020}
{Minniti}, J.~H., {Sbordone}, L., {Rojas-Arriagada}, A., {et~al.} 2020, \aap,
  640, A92

\bibitem[{{Minniti} {et~al.}(2021{\natexlab{b}}){Minniti}, {Zoccali},
  {Rojas-Arriagada}, {Minniti}, {Sbordone}, {Contreras Ramos}, {Braga},
  {Catelan}, {Duffau}, {Gieren}, {Marconi}, \& {Valcarce}}]{jminniti2021}
{Minniti}, J.~H., {Zoccali}, M., {Rojas-Arriagada}, A., {et~al.}
  2021{\natexlab{b}}, \aap, 654, A138

\bibitem[{{Murali} \& {Weinberg}(1997)}]{murali1997}
{Murali}, C. \& {Weinberg}, M.~D. 1997, \mnras, 288, 749

\bibitem[{{Myeong} {et~al.}(2019){Myeong}, {Vasiliev}, {Iorio}, {Evans}, \&
  {Belokurov}}]{myeong2019}
{Myeong}, G.~C., {Vasiliev}, E., {Iorio}, G., {Evans}, N.~W., \& {Belokurov},
  V. 2019, \mnras, 488, 1235

\bibitem[{{Naidu} {et~al.}(2020){Naidu}, {Conroy}, {Bonaca}, {Johnson}, {Ting},
  {Caldwell}, {Zaritsky}, \& {Cargile}}]{naidu2020}
{Naidu}, R.~P., {Conroy}, C., {Bonaca}, A., {et~al.} 2020, \apj, 901, 48

\bibitem[{{Ness} {et~al.}(2015){Ness}, {Hogg}, {Rix}, {Ho}, \&
  {Zasowski}}]{ness2015}
{Ness}, M., {Hogg}, D.~W., {Rix}, H.~W., {Ho}, A. Y.~Q., \& {Zasowski}, G.
  2015, \apj, 808, 16

\bibitem[{Nidever(2021)}]{nidever2021}
Nidever, D. 2021, dnidever/doppler: Cannon and Payne models

\bibitem[{{Ochsenbein} {et~al.}(2000){Ochsenbein}, {Bauer}, \&
  {Marcout}}]{ochsenbein2000}
{Ochsenbein}, F., {Bauer}, P., \& {Marcout}, J. 2000, \aaps, 143, 23

\bibitem[{{Pace} {et~al.}(2023){Pace}, {Koposov}, {Walker}, {Caldwell},
  {Mateo}, {Olszewski}, {Roederer}, {Bailey}, {Belokurov}, {Kuehn}, {Li}, \&
  {Zucker}}]{pace2023}
{Pace}, A.~B., {Koposov}, S.~E., {Walker}, M.~G., {et~al.} 2023, arXiv
  e-prints, arXiv:2304.06904

\bibitem[{{Pagnini} {et~al.}(2022){Pagnini}, {Di Matteo}, {Khoperskov},
  {Mastrobuono-Battisti}, {Haywood}, {Renaud}, \& {Combes}}]{pagnini2022}
{Pagnini}, G., {Di Matteo}, P., {Khoperskov}, S., {et~al.} 2022, arXiv
  e-prints, arXiv:2210.04245

\bibitem[{P\'erez \& Granger(2007)}]{perez2007}
P\'erez, F. \& Granger, B.~E. 2007, Computing in Science and Engineering, 9, 21

\bibitem[{Price-Whelan {et~al.}(2020)Price-Whelan, Sipőcz, Lenz, Greco,
  Starkman, Foreman-Mackey, Lim, Oh, Koposov, \& Major}]{pricewhelan2020}
Price-Whelan, A., Sipőcz, B., Lenz, D., {et~al.} 2020, adrn/gala: v1.3

\bibitem[{Price-Whelan(2017)}]{pricewhelan2017}
Price-Whelan, A.~M. 2017, The Journal of Open Source Software, 2

\bibitem[{{Recio-Blanco}(2018)}]{recioblanco2018}
{Recio-Blanco}, A. 2018, \aap, 620, A194

\bibitem[{{Renzini}(2017)}]{renzini2017}
{Renzini}, A. 2017, \mnras, 469, L63

\bibitem[{{Rojas-Arriagada} {et~al.}(2020){Rojas-Arriagada}, {Zasowski},
  {Schultheis}, {Zoccali}, {Hasselquist}, {Chiappini}, {Cohen}, {Cunha},
  {Fern{\'a}ndez-Trincado}, {Fragkoudi}, {Garc{\'\i}a-Hern{\'a}ndez},
  {Geisler}, {Gran}, {Lian}, {Majewski}, {Minniti}, {Monachesi}, {Nitschelm},
  \& {Queiroz}}]{rojasarriagada2020}
{Rojas-Arriagada}, A., {Zasowski}, G., {Schultheis}, M., {et~al.} 2020, \mnras,
  499, 1037

\bibitem[{{Ryu} \& {Lee}(2018)}]{ryu2018}
{Ryu}, J. \& {Lee}, M.~G. 2018, \apjl, 863, L38

\bibitem[{{Saito} {et~al.}(2020){Saito}, {Minniti}, {Benjamin}, {Navarro},
  {Alonso-Garc{\'\i}a}, {Gonzalez}, {Kammers}, \& {Surot}}]{saito2020}
{Saito}, R.~K., {Minniti}, D., {Benjamin}, R.~A., {et~al.} 2020, \mnras, 494,
  L32

\bibitem[{{Sanna} {et~al.}(2017){Sanna}, {Reid}, {Dame}, {Menten}, \&
  {Brunthaler}}]{sanna2017}
{Sanna}, A., {Reid}, M.~J., {Dame}, T.~M., {Menten}, K.~M., \& {Brunthaler}, A.
  2017, Science, 358, 227

\bibitem[{{Smith} {et~al.}(2018){Smith}, {Lucas}, {Kurtev}, {Smart}, {Minniti},
  {Borissova}, {Jones}, {Zhang}, {Marocco}, {Contreras Pe{\~n}a}, {Gromadzki},
  {Kuhn}, {Drew}, {Pinfield}, \& {Bedin}}]{virac}
{Smith}, L.~C., {Lucas}, P.~W., {Kurtev}, R., {et~al.} 2018, \mnras, 474, 1826

\bibitem[{{Str{\"o}bele} {et~al.}(2012){Str{\"o}bele}, {La Penna}, {Arsenault},
  {Conzelmann}, {Delabre}, {Duchateau}, {Dorn}, {Fedrigo}, {Hubin}, {Quentin},
  {Jolley}, {Kiekebusch}, {Kirchbauer}, {Klein}, {Kolb}, {Kuntschner}, {Le
  Louarn}, {Lizon}, {Madec}, {Pettazzi}, {Soenke}, {Tordo}, {Vernet}, \&
  {Muradore}}]{strobele2012}
{Str{\"o}bele}, S., {La Penna}, P., {Arsenault}, R., {et~al.} 2012, in Society
  of Photo-Optical Instrumentation Engineers (SPIE) Conference Series, Vol.
  8447, Adaptive Optics Systems III, ed. B.~L. {Ellerbroek}, E.~{Marchetti}, \&
  J.-P. {V{\'e}ran}, 844737

\bibitem[{{Stuik} {et~al.}(2006){Stuik}, {Bacon}, {Conzelmann}, {Delabre},
  {Fedrigo}, {Hubin}, {Le Louarn}, \& {Str{\"o}bele}}]{stuik2006}
{Stuik}, R., {Bacon}, R., {Conzelmann}, R., {et~al.} 2006, \nar, 49, 618

\bibitem[{{Surot} {et~al.}(2019){Surot}, {Valenti}, {Hidalgo}, {Zoccali},
  {S{\"o}kmen}, {Rejkuba}, {Minniti}, {Gonzalez}, {Cassisi}, {Renzini}, \&
  {Weiss}}]{surot2019}
{Surot}, F., {Valenti}, E., {Hidalgo}, S.~L., {et~al.} 2019, \aap, 623, A168

\bibitem[{{Taylor}(2005)}]{topcat}
{Taylor}, M.~B. 2005, in Astronomical Society of the Pacific Conference Series,
  Vol. 347, Astronomical Data Analysis Software and Systems XIV, ed.
  P.~{Shopbell}, M.~{Britton}, \& R.~{Ebert}, 29

\bibitem[{{Vanzella} {et~al.}(2017){Vanzella}, {Calura}, {Meneghetti},
  {Mercurio}, {Castellano}, {Caminha}, {Balestra}, {Rosati}, {Tozzi}, {De
  Barros}, {Grazian}, {D'Ercole}, {Ciotti}, {Caputi}, {Grillo}, {Merlin},
  {Pentericci}, {Fontana}, {Cristiani}, \& {Coe}}]{vanzella2017}
{Vanzella}, E., {Calura}, F., {Meneghetti}, M., {et~al.} 2017, \mnras, 467,
  4304

\bibitem[{{Vasiliev} \& {Baumgardt}(2021)}]{vasiliev2021}
{Vasiliev}, E. \& {Baumgardt}, H. 2021, \mnras, 505, 5978

\bibitem[{{Wang} {et~al.}(2022){Wang}, {Hayden}, {Sharma}, {Xiang}, {Ting},
  {Bland-Hawthorn}, \& {Chen}}]{wang2022}
{Wang}, Z., {Hayden}, M.~R., {Sharma}, S., {et~al.} 2022, \mnras, 514, 1034

\bibitem[{{Watkins} {et~al.}(2015){Watkins}, {van der Marel}, {Bellini}, \&
  {Anderson}}]{watkins2015}
{Watkins}, L.~L., {van der Marel}, R.~P., {Bellini}, A., \& {Anderson}, J.
  2015, \apj, 803, 29

\bibitem[{{Webb} \& {Carlberg}(2021)}]{webb2021}
{Webb}, J.~J. \& {Carlberg}, R.~G. 2021, \mnras, 502, 4547

\bibitem[{{Weilbacher} {et~al.}(2020){Weilbacher}, {Palsa}, {Streicher},
  {Bacon}, {Urrutia}, {Wisotzki}, {Conseil}, {Husemann}, {Jarno}, {Kelz},
  {P{\'e}contal-Rousset}, {Richard}, {Roth}, {Selman}, \&
  {Vernet}}]{weilbacher2020}
{Weilbacher}, P.~M., {Palsa}, R., {Streicher}, O., {et~al.} 2020, \aap, 641,
  A28

\end{thebibliography}

\begin{appendix} %First appendix
\section{Appendix: additional GC plots and complete membership catalogues}
\label{sec:appendix}
% Explanation of the purpose of the appendix: 

\begin{figure*}
\centering
\includegraphics[scale=0.5]{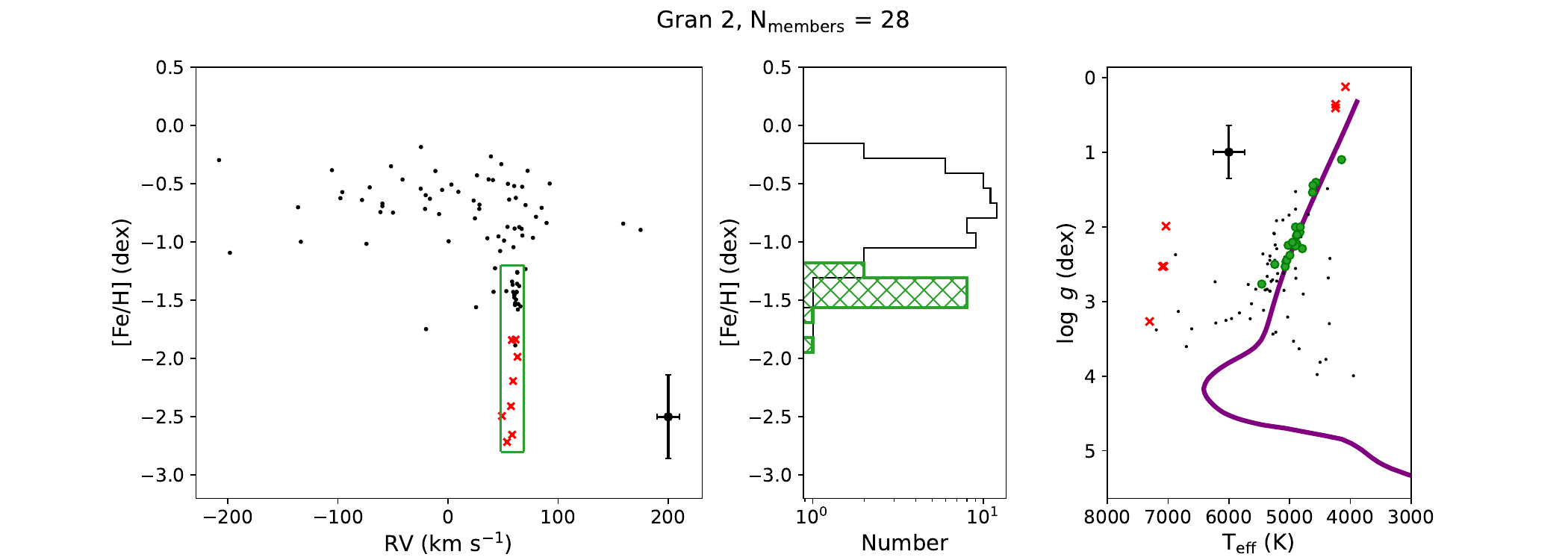}\\
\includegraphics[scale=0.5]{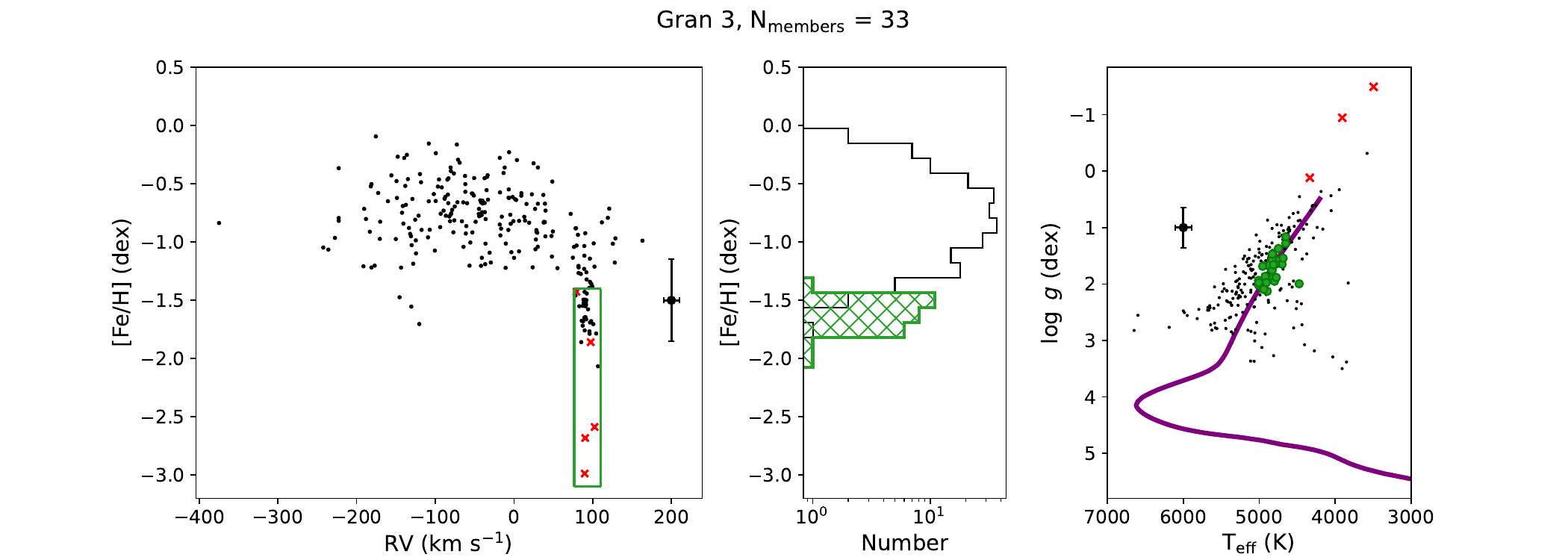}\\
\includegraphics[scale=0.5]{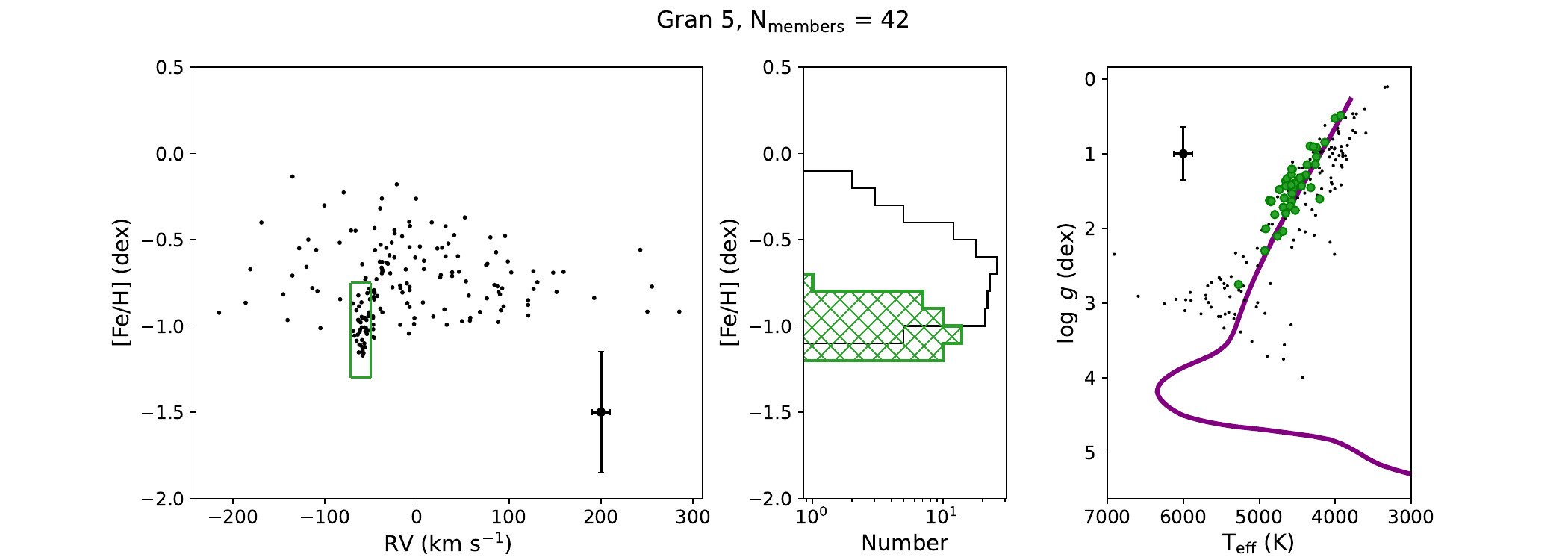}
\caption{{\bf Left panels}: RV-[Fe/H] plane for all the MUSE extracted stars in the Gran~2 ({\bf upper row}), Gran~3 ({\bf middle row}) and Gran~5 ({\bf lower row}) fields. 
The box was drawn to select the cluster members using the individual RV and metallicity values. 
Red crosses represent stars outside the calibration parameter space, i.e., stars in the RGB tip or HB/SGB stars, 
whose RV coincides with that derived for the cluster, but with no reliable stellar atmospheric parameter determination.
{\bf Middle panels}: Metallicity histogram for cluster selected stars. 
Highlighted hatched-filled histogram contains the cluster members within the box. 
{\bf Right panels}: Kiel diagram for the field and cluster stars in grey and green, respectively. 
A PARSEC isochrone was drawn using the same derived metallicity with an age of 10 Gyr.
Number of members considers all the valid RV cluster stars, i.e., members within the highlighted box.
The errorbar located in each panel represents the mean uncertainty for each corresponding parameter.}
\label{fig:triplots}
\end{figure*}

\begin{table*}
\caption{Atmospheric parameters for all the MUSE Gran~1 analysed stars.
Star ID, coordinates, temperature, surface gravity, metallicity, RV, and mean SNR were derived for observed GC members based on their multidimensional membership (RV, [Fe/H], and Kiel diagram).}
\centering
\label{tbl:allstars1}
\begin{tabular}{c c c c c c c c}
\hline\hline
GC - ID & RA & Dec & $T_{\rm eff}$ & $\log\ g$ & ${\rm [Fe/H]}$ & RV & SNR\\
        & (deg) & (deg) & (K) & (dex) & (dex) & (km s$^{-1}$) & \\
\hline
Gran 01 - 008 & 269.64794 & -32.02559 & 4362 & 1.2 & -1.17 & 68.25 & 133 \\
Gran 01 - 017 & 269.64650 & -32.02449 & 4659 & 1.4 & -1.11 & 69.33 & 174 \\
Gran 01 - 021 & 269.65005 & -32.02420 & 4358 & 1.1 & -1.14 & 75.04 & 241 \\
Gran 01 - 022 & 269.65615 & -32.02395 & 4611 & 1.4 & -1.05 & 68.81 & 222 \\
Gran 01 - 027 & 269.66006 & -32.02366 & 4577 & 1.2 & -1.08 & 78.36 & 177 \\
Gran 01 - 028 & 269.65463 & -32.02359 & 4521 & 1.2 & -1.17 & 76.71 & 185 \\
Gran 01 - 031 & 269.65405 & -32.02354 & 4416 & 1.2 & -1.13 & 75.47 & 119 \\
Gran 01 - 033 & 269.65832 & -32.02317 & ---- & --- & ----- & 80.74 & 371 \\
Gran 01 - 034 & 269.65652 & -32.02305 & 4540 & 1.1 & -1.19 & 76.46 & 167 \\
Gran 01 - 040 & 269.65639 & -32.02255 & 4292 & 0.7 & -1.26 & 74.18 & 64 \\
Gran 01 - 041 & 269.65260 & -32.02249 & 4153 & 0.7 & -1.15 & 78.03 & 302 \\
Gran 01 - 042 & 269.64663 & -32.02239 & 4353 & 1.2 & -1.13 & 77.38 & 121 \\
Gran 01 - 044 & 269.64840 & -32.02222 & 4627 & 1.5 & -1.10 & 80.21 & 101 \\
Gran 01 - 048 & 269.65071 & -32.02198 & 4212 & 0.9 & -1.12 & 81.96 & 293 \\
Gran 01 - 056 & 269.65463 & -32.02156 & 4131 & 0.8 & -1.16 & 77.37 & 232 \\
Gran 01 - 059 & 269.64769 & -32.02121 & 4757 & 1.4 & -1.02 & 82.67 & 168 \\
Gran 01 - 061 & 269.64709 & -32.02105 & 4776 & 1.8 & -1.02 & 77.03 & 79 \\
Gran 01 - 065 & 269.64866 & -32.02094 & 4435 & 1.1 & -1.06 & 77.35 & 106 \\
Gran 01 - 066 & 269.65901 & -32.02082 & 4245 & 0.7 & -1.12 & 80.70 & 285 \\
Gran 01 - 076 & 269.64899 & -32.02021 & 4251 & 0.9 & -1.09 & 72.58 & 198 \\
Gran 01 - 082 & 269.66191 & -32.01964 & 4572 & 1.5 & -1.17 & 73.48 & 43 \\
Gran 01 - 083 & 269.65417 & -32.01947 & 3994 & 0.6 & -1.17 & 77.06 & 313 \\
Gran 01 - 087 & 269.64881 & -32.01904 & 4477 & 1.4 & -1.01 & 78.37 & 151 \\
Gran 01 - 089 & 269.64552 & -32.01901 & 4371 & 1.1 & -1.15 & 78.18 & 139 \\
Gran 01 - 090 & 269.65437 & -32.01884 & 4358 & 1.1 & -1.21 & 75.96 & 142 \\
Gran 01 - 091 & 269.64879 & -32.01884 & 4501 & 1.5 & -1.01 & 81.11 & 143 \\
Gran 01 - 095 & 269.65619 & -32.01849 & 4447 & 1.2 & -1.20 & 85.42 & 102 \\
Gran 01 - 098 & 269.64272 & -32.01788 & 4260 & 1.1 & -1.09 & 78.35 & 80 \\
Gran 01 - 102 & 269.64590 & -32.01726 & 4627 & 1.3 & -1.16 & 74.37 & 189 \\
Gran 01 - 103 & 269.65461 & -32.01716 & 4611 & 1.7 & -1.15 & 78.97 & 83 \\
Gran 01 - 104 & 269.65095 & -32.01700 & 4233 & 0.9 & -1.06 & 71.73 & 322 \\
Gran 01 - 106 & 269.65882 & -32.01693 & 4019 & 0.7 & -1.13 & 77.90 & 267 \\
Gran 01 - 107 & 269.65583 & -32.01694 & 4481 & 1.3 & -1.14 & 75.62 & 96 \\
Gran 01 - 108 & 269.65208 & -32.01644 & 4515 & 1.3 & -1.08 & 74.41 & 163 \\
Gran 01 - 113 & 269.65306 & -32.01586 & 4253 & 0.9 & -1.21 & 78.18 & 136 \\
Gran 01 - 118 & 269.65076 & -32.01565 & 4664 & 1.3 & -1.08 & 75.08 & 161 \\
Gran 01 - 126 & 269.65174 & -32.01465 & 4378 & 1.2 & -1.12 & 77.83 & 129 \\
Gran 01 - 128 & 269.64577 & -32.01443 & 4517 & 1.4 & -1.17 & 72.40 & 93 \\
Gran 01 - 129 & 269.64302 & -32.01433 & 4282 & 1.1 & -1.12 & 78.51 & 180 \\
Gran 01 - 132 & 269.64382 & -32.01395 & 4312 & 1.0 & -1.25 & 80.51 & 141 \\
Gran 01 - 140 & 269.65175 & -32.01283 & 4184 & 0.8 & -1.14 & 75.95 & 301 \\
Gran 01 - 147 & 269.65825 & -32.01144 & 4262 & 0.9 & -1.11 & 81.18 & 191 \\
Gran 01 - 151 & 269.64231 & -32.01100 & 4584 & 1.1 & -1.12 & 81.10 & 88 \\
\hline
\end{tabular}
\end{table*}

\begin{table*}
\caption{Atmospheric parameters for all the MUSE Gran~2 analysed stars.
Star ID, coordinates, temperature, surface gravity, metallicity, RV, and mean SNR were derived for observed GC members based on their multidimensional membership (RV, [Fe/H], and Kiel diagram).}
\centering
\label{tbl:allstars2}
\begin{tabular}{c c c c c c c c}
\hline\hline
GC - ID & RA & Dec & $T_{\rm eff}$ & $\log\ g$ & ${\rm [Fe/H]}$ & RV & SNR\\
        & (deg) & (deg) & (K) & (dex) & (dex) & (km s$^{-1}$) & \\
\hline
Gran 02 - 004 & 257.89591 & -24.85627 & ---- & --- & ----- & 61.41 & 446 \\
Gran 02 - 006 & 257.89126 & -24.85582 & ---- & --- & ----- & 57.91 & 357 \\
Gran 02 - 016 & 257.89701 & -24.85381 & 4903 & 2.0 & -1.48 & 60.22 & 127 \\
Gran 02 - 019 & 257.89022 & -24.85292 & 4791 & 2.3 & -1.42 & 52.90 & 250 \\
Gran 02 - 021 & 257.89498 & -24.85237 & 4883 & 2.2 & -1.58 & 63.43 & 146 \\
Gran 02 - 025 & 257.89939 & -24.85186 & 5024 & 2.2 & -1.55 & 65.79 & 146 \\
Gran 02 - 026 & 257.89908 & -24.85177 & 5244 & 2.5 & -1.53 & 63.77 & 104 \\
Gran 02 - 028 & 257.89909 & -24.85161 & 5459 & 2.8 & -1.36 & 62.70 & 82 \\
Gran 02 - 034 & 257.89351 & -24.85104 & 4887 & 2.1 & -1.54 & 60.77 & 123 \\
Gran 02 - 035 & 257.89755 & -24.85093 & 5060 & 2.5 & -1.34 & 58.15 & 67 \\
Gran 02 - 036 & 257.89315 & -24.85064 & 4905 & 2.3 & -1.38 & 64.63 & 67 \\
Gran 02 - 038 & 257.88808 & -24.85059 & ---- & --- & ----- & 58.22 & 103 \\
Gran 02 - 041 & 257.89621 & -24.85003 & ---- & --- & ----- & 48.96 & 70 \\
Gran 02 - 052 & 257.88781 & -24.84909 & ---- & --- & ----- & 57.16 & 221 \\
Gran 02 - 053 & 257.88389 & -24.84914 & ---- & --- & ----- & 53.58 & 94 \\
Gran 02 - 055 & 257.88941 & -24.84859 & 5074 & 2.5 & -1.26 & 62.84 & 53 \\
Gran 02 - 060 & 257.89506 & -24.84777 & 4831 & 2.1 & -1.52 & 61.00 & 145 \\
Gran 02 - 061 & 257.88843 & -24.84776 & 4826 & 2.0 & -1.37 & 58.69 & 65 \\
Gran 02 - 063 & 257.89940 & -24.84760 & 4569 & 1.4 & -1.49 & 61.72 & 311 \\
Gran 02 - 065 & 257.89433 & -24.84742 & ---- & --- & ----- & 59.09 & 110 \\
Gran 02 - 074 & 257.89690 & -24.84643 & 4616 & 1.4 & -1.43 & 62.19 & 263 \\
Gran 02 - 076 & 257.89143 & -24.84603 & 4953 & 2.2 & -1.43 & 59.16 & 121 \\
Gran 02 - 088 & 257.88727 & -24.84371 & 4623 & 1.5 & -1.43 & 62.34 & 248 \\
Gran 02 - 092 & 257.88665 & -24.84299 & 4879 & 2.1 & -1.45 & 60.43 & 112 \\
Gran 02 - 093 & 257.89169 & -24.84255 & 4146 & 1.1 & -1.89 & 61.02 & 285 \\
Gran 02 - 094 & 257.88905 & -24.84238 & 5047 & 2.4 & -1.46 & 60.14 & 139 \\
Gran 02 - 095 & 257.89346 & -24.84232 & 4997 & 2.4 & -1.26 & 62.89 & 54 \\
Gran 02 - 103 & 257.88298 & -24.84043 & ---- & --- & ----- & 63.04 & 299 \\
\hline
\end{tabular}
\end{table*}

\begin{table*}
\caption{Atmospheric parameters for all the MUSE Gran~3 analysed stars.
Star ID, coordinates, temperature, surface gravity, metallicity, RV, and mean SNR were derived for observed GC members based on their multidimensional membership (RV, [Fe/H], and Kiel diagram).}
\centering
\label{tbl:allstars3}
\begin{tabular}{c c c c c c c c}
\hline\hline
GC - ID & RA & Dec & $T_{\rm eff}$ & $\log\ g$ & ${\rm [Fe/H]}$ & RV & SNR\\
        & (deg) & (deg) & (K) & (dex) & (dex) & (km s$^{-1}$) & \\
\hline
Gran 03 - 032 & 256.26735 & -35.49913 & 4818 & 1.5 & -1.52 & 91.34 & 73 \\
Gran 03 - 034 & 256.25288 & -35.49906 & 4822 & 1.9 & -1.68 & 98.00 & 53 \\
Gran 03 - 074 & 256.26345 & -35.49795 & 4766 & 1.6 & -1.50 & 91.92 & 71 \\
Gran 03 - 091 & 256.25422 & -35.49756 & 4850 & 1.7 & -1.55 & 87.47 & 83 \\
Gran 03 - 101 & 256.24863 & -35.49737 & ---- & ---- & ---- & 90.49 & 214 \\
Gran 03 - 152 & 256.25697 & -35.49551 & 4868 & 1.7 & -1.50 & 88.16 & 66 \\
Gran 03 - 163 & 256.25260 & -35.49513 & 4922 & 1.9 & -1.65 & 89.46 & 111 \\
Gran 03 - 169 & 256.26344 & -35.49499 & 4949 & 2.1 & -1.72 & 87.38 & 69 \\
Gran 03 - 176 & 256.26352 & -35.49485 & 4894 & 2.1 & -1.86 & 85.48 & 76 \\
Gran 03 - 189 & 256.26611 & -35.49458 & 4831 & 1.6 & -1.68 & 86.04 & 88 \\
Gran 03 - 198 & 256.26460 & -35.49447 & 4826 & 1.5 & -1.76 & 89.79 & 109 \\
Gran 03 - 204 & 256.25689 & -35.49441 & 4800 & 2.0 & -1.54 & 91.51 & 56 \\
Gran 03 - 302 & 256.25186 & -35.49234 & 4829 & 1.8 & -1.58 & 94.05 & 67 \\
Gran 03 - 334 & 256.24962 & -35.49163 & 4698 & 1.6 & -1.52 & 88.57 & 57 \\
Gran 03 - 357 & 256.25547 & -35.49105 & 4652 & 1.3 & -1.68 & 87.76 & 118 \\
Gran 03 - 362 & 256.25493 & -35.49092 & ---- & ---- & ---- & 89.90 & 48 \\
Gran 03 - 367 & 256.25521 & -35.49079 & 4776 & 1.9 & -1.49 & 89.72 & 52 \\
Gran 03 - 397 & 256.26641 & -35.49015 & 5010 & 2.0 & -1.46 & 78.17 & 219 \\
Gran 03 - 417 & 256.25740 & -35.48946 & 4788 & 1.7 & -1.79 & 96.10 & 114 \\
Gran 03 - 448 & 256.25610 & -35.48885 & 5009 & 1.9 & -1.55 & 90.80 & 99 \\
Gran 03 - 479 & 256.26517 & -35.48813 & 4475 & 2.0 & -2.07 & 106.77 & 68 \\
Gran 03 - 485 & 256.25616 & -35.48791 & 4748 & 1.4 & -1.44 & 92.44 & 204 \\
Gran 03 - 519 & 256.26526 & -35.48692 & ---- & --- & ----- & 79.33 & 211 \\
Gran 03 - 533 & 256.25561 & -35.48652 & 4955 & 1.7 & -1.55 & 83.07 & 122 \\
Gran 03 - 542 & 256.25971 & -35.48608 & 4814 & 1.7 & -1.69 & 96.64 & 111 \\
Gran 03 - 551 & 256.26722 & -35.48564 & ---- & --- & ----- & 102.42 & 24 \\
Gran 03 - 563 & 256.26245 & -35.48517 & 4941 & 2.1 & -1.70 & 100.79 & 111 \\
Gran 03 - 567 & 256.26230 & -35.48501 & 4907 & 2.0 & -1.66 & 91.72 & 124 \\
Gran 03 - 569 & 256.26380 & -35.48497 & 4827 & 1.5 & -1.43 & 89.73 & 100 \\
Gran 03 - 584 & 256.25732 & -35.48457 & ---- & --- & ----- & 97.47 & 123 \\
Gran 03 - 592 & 256.26850 & -35.48423 & 4650 & 1.2 & -1.64 & 90.75 & 188 \\
Gran 03 - 600 & 256.25726 & -35.48407 & 5006 & 2.0 & -1.79 & 104.42 & 139 \\
Gran 03 - 605 & 256.25759 & -35.48387 & 4686 & 1.5 & -1.77 & 95.87 & 51 \\
\hline
\end{tabular}
\end{table*}

\begin{table*}
\caption{Atmospheric parameters for all the MUSE Gran~4 analysed stars.
Star ID, coordinates, temperature, surface gravity, metallicity, RV, and mean SNR were derived for observed GC members based on their multidimensional membership (RV, [Fe/H], and Kiel diagram).}
\centering
\label{tbl:allstars4}
\begin{tabular}{c c c c c c c c}
\hline\hline
GC - ID & RA & Dec & $T_{\rm eff}$ & $\log\ g$ & ${\rm [Fe/H]}$ & RV & SNR\\
        & (deg) & (deg) & (K) & (dex) & (dex) & (km s$^{-1}$) & \\
\hline
Gran 04 - 030 & 278.10913 & -23.11954 & ---- & --- & ----- & -264.43 & 137 \\
Gran 04 - 105 & 278.10611 & -23.11620 & ---- & --- & ----- & -267.35 & 136 \\
Gran 04 - 119 & 278.11364 & -23.11583 & ---- & --- & ----- & -267.16 & 133 \\
Gran 04 - 125 & 278.12205 & -23.11561 & 5016 & 1.9 & -1.93 & -267.32 & 168 \\
Gran 04 - 131 & 278.11607 & -23.11537 & 4818 & 1.9 & -1.94 & -263.74 & 230 \\
Gran 04 - 133 & 278.12187 & -23.11533 & 4819 & 2.1 & -1.55 & -269.68 & 65 \\
Gran 04 - 138 & 278.11595 & -23.11512 & 5107 & 2.8 & -1.71 & -262.05 & 179 \\
Gran 04 - 155 & 278.11657 & -23.11421 & ---- & --- & ----- & -259.20 & 144 \\
Gran 04 - 163 & 278.11162 & -23.11389 & ---- & --- & ----- & -269.37 & 129 \\
Gran 04 - 166 & 278.11517 & -23.11376 & 4743 & 2.2 & -2.00 & -264.95 & 231 \\
Gran 04 - 169 & 278.11323 & -23.11371 & ---- & --- & ----- & -262.28 & 29 \\
Gran 04 - 173 & 278.11554 & -23.11356 & 4531 & 0.9 & -1.24 & -272.67 & 45 \\
Gran 04 - 205 & 278.12211 & -23.11244 & ---- & --- & ----- & -264.56 & 27 \\
Gran 04 - 228 & 278.11179 & -23.11137 & 5437 & 3.0 & -1.85 & -257.95 & 78 \\
Gran 04 - 238 & 278.12152 & -23.11100 & 5458 & 2.9 & -1.12 & -264.80 & 50 \\
Gran 04 - 246 & 278.12139 & -23.11054 & 5381 & 2.9 & -1.48 & -262.93 & 65 \\
Gran 04 - 250 & 278.11547 & -23.11050 & ---- & --- & ----- & -266.67 & 136 \\
Gran 04 - 293 & 278.10536 & -23.10850 & ---- & --- & ----- & -276.15 & 143 \\
Gran 04 - 296 & 278.12011 & -23.10823 & 5122 & 2.0 & -0.92 & -270.83 & 34 \\
Gran 04 - 312 & 278.11662 & -23.10762 & 5316 & 3.0 & -1.57 & -260.41 & 74 \\
Gran 04 - 324 & 278.10822 & -23.10686 & 5033 & 2.3 & -1.31 & -262.18 & 45 \\
Gran 04 - 326 & 278.11352 & -23.10683 & 5230 & 2.7 & -1.74 & -264.10 & 119 \\
Gran 04 - 339 & 278.11269 & -23.10611 & ---- & --- & ----- & -269.44 & 26 \\
Gran 04 - 344 & 278.11325 & -23.10594 & 4947 & 2.0 & -1.96 & -266.81 & 203 \\
Gran 04 - 349 & 278.10683 & -23.10553 & 5193 & 2.5 & -1.82 & -270.24 & 119 \\
Gran 04 - 368 & 278.10652 & -23.10461 & 5142 & 2.8 & -1.78 & -263.75 & 79 \\
\hline
\end{tabular}
\end{table*}

\begin{table*}
\caption{Atmospheric parameters for all the MUSE Gran~5 analysed stars.
Star ID, coordinates, temperature, surface gravity, metallicity, RV, and mean SNR were derived for observed GC members based on their multidimensional membership (RV, [Fe/H], and Kiel diagram).}
\centering
\label{tbl:allstars5}
\begin{tabular}{c c c c c c c c}
\hline\hline
GC - ID & RA & Dec & $T_{\rm eff}$ & $\log\ g$ & ${\rm [Fe/H]}$ & RV & SNR\\
        & (deg) & (deg) & (K) & (dex) & (dex) & (km s$^{-1}$) & \\
\hline
Gran 05 - 027 & 267.22633 & -24.17476 & 4846 & 1.6 & -0.96 & -63.81 & 236 \\
Gran 05 - 029 & 267.22775 & -24.17466 & 4915 & 2.0 & -0.96 & -58.90 & 68 \\
Gran 05 - 035 & 267.22945 & -24.17438 & 4136 & 0.8 & -0.87 & -69.43 & 134 \\
Gran 05 - 036 & 267.22548 & -24.17431 & 4865 & 1.6 & -0.89 & -62.80 & 281 \\
Gran 05 - 042 & 267.22750 & -24.17393 & 4248 & 0.9 & -1.12 & -59.38 & 247 \\
Gran 05 - 044 & 267.22895 & -24.17381 & 4444 & 1.4 & -1.08 & -59.68 & 185 \\
Gran 05 - 049 & 267.22554 & -24.17336 & 4652 & 1.4 & -1.02 & -61.75 & 269 \\
Gran 05 - 057 & 267.23091 & -24.17259 & 4563 & 1.2 & -1.03 & -62.12 & 233 \\
Gran 05 - 058 & 267.22522 & -24.17259 & 4249 & 1.0 & -0.85 & -51.03 & 55 \\
Gran 05 - 061 & 267.22553 & -24.17243 & 4534 & 1.4 & -1.10 & -57.10 & 102 \\
Gran 05 - 067 & 267.22840 & -24.17209 & 4388 & 1.3 & -1.05 & -64.24 & 64 \\
Gran 05 - 068 & 267.23760 & -24.17204 & 4369 & 1.1 & -1.01 & -59.89 & 232 \\
Gran 05 - 069 & 267.23280 & -24.17170 & 4684 & 1.7 & -0.81 & -50.32 & 44 \\
Gran 05 - 073 & 267.22329 & -24.17153 & 4634 & 1.3 & -0.97 & -54.73 & 347 \\
Gran 05 - 078 & 267.22841 & -24.17116 & 4003 & 0.5 & -1.05 & -54.28 & 474 \\
Gran 05 - 081 & 267.23992 & -24.17094 & 4261 & 1.1 & -0.94 & -52.70 & 80 \\
Gran 05 - 084 & 267.22811 & -24.17060 & 4454 & 1.3 & -1.03 & -56.04 & 76 \\
Gran 05 - 088 & 267.23146 & -24.17020 & 4573 & 1.6 & -1.11 & -62.53 & 110 \\
Gran 05 - 089 & 267.22906 & -24.17015 & 4763 & 2.1 & -1.12 & -56.42 & 113 \\
Gran 05 - 100 & 267.22622 & -24.16938 & 4330 & 0.9 & -1.15 & -59.29 & 250 \\
Gran 05 - 104 & 267.22273 & -24.16927 & 4927 & 2.3 & -0.82 & -63.61 & 121 \\
Gran 05 - 106 & 267.23644 & -24.16894 & 4735 & 1.5 & -0.98 & -61.20 & 226 \\
Gran 05 - 110 & 267.22883 & -24.16858 & 4688 & 2.0 & -1.01 & -57.12 & 87 \\
Gran 05 - 112 & 267.23578 & -24.16848 & 4555 & 1.5 & -1.11 & -61.57 & 128 \\
Gran 05 - 116 & 267.23394 & -24.16831 & 4287 & 0.9 & -1.15 & -63.20 & 254 \\
Gran 05 - 119 & 267.22347 & -24.16813 & 4568 & 1.5 & -1.16 & -57.94 & 103 \\
Gran 05 - 120 & 267.23420 & -24.16809 & 4672 & 1.6 & -1.03 & -51.75 & 166 \\
Gran 05 - 129 & 267.22876 & -24.16765 & 4652 & 1.8 & -1.03 & -69.44 & 124 \\
Gran 05 - 131 & 267.22614 & -24.16760 & 4575 & 1.3 & -1.01 & -55.48 & 55 \\
Gran 05 - 137 & 267.23877 & -24.16681 & 4842 & 1.6 & -0.95 & -59.93 & 232 \\
Gran 05 - 156 & 267.23577 & -24.16442 & 4573 & 1.2 & -0.95 & -56.62 & 367 \\
Gran 05 - 160 & 267.22267 & -24.16432 & 4595 & 1.7 & -1.02 & -53.59 & 144 \\
Gran 05 - 161 & 267.23078 & -24.16419 & 4535 & 1.4 & -0.92 & -52.07 & 284 \\
Gran 05 - 162 & 267.22791 & -24.16421 & 4652 & 1.4 & -0.97 & -63.20 & 240 \\
Gran 05 - 167 & 267.22864 & -24.16359 & 4796 & 1.8 & -1.06 & -67.82 & 66 \\
Gran 05 - 176 & 267.22931 & -24.16247 & 4466 & 1.3 & -0.91 & -65.73 & 41 \\
Gran 05 - 177 & 267.22653 & -24.16249 & 4204 & 1.6 & -0.81 & -53.90 & 103 \\
Gran 05 - 178 & 267.22700 & -24.16242 & 4321 & 1.5 & -0.86 & -60.25 & 226 \\
Gran 05 - 179 & 267.23139 & -24.16237 & 4581 & 1.4 & -1.09 & -65.49 & 53 \\
Gran 05 - 180 & 267.22852 & -24.16232 & 5272 & 2.8 & -0.78 & -51.57 & 147 \\
Gran 05 - 183 & 267.22476 & -24.16209 & 4527 & 1.8 & -1.17 & -58.93 & 141 \\
Gran 05 - 188 & 267.22261 & -24.16170 & 3931 & 0.5 & -1.14 & -58.94 & 422 \\
\hline
\end{tabular}
\end{table*}

\end{appendix}

\end{document}